\documentclass[journal=jctcce,manuscript=article,layout=twocolumn]{achemso} 

\usepackage{graphicx}
\usepackage{gensymb}
\usepackage[aboveskip=0pt,belowskip=-5pt]{caption}
\usepackage{subfig}
\usepackage{dcolumn}
\usepackage{bm}
\usepackage[version=4]{mhchem}
\usepackage{braket}
\usepackage[export]{adjustbox}
\usepackage[section]{placeins}
\usepackage{multirow}
\usepackage[utf8]{inputenc}
\usepackage[T1]{fontenc}
\usepackage{threeparttable}
\setkeys{acs}{articletitle=true}
\graphicspath{{./}}
\title[]{Multireference Stochastic Coupled Cluster}

\author{Maria-Andreea Filip}
 \email{maf63@cam.ac.uk}
\author{Charles J C Scott}
\author{Alex J W Thom}
 \email{ajwt3@cam.ac.uk}

\affiliation{ 
Department of Chemistry, University of Cambridge, Cambridge, UK
}%

\begin{document}

\begin{abstract}

We describe a modification of the stochastic coupled cluster algorithm that 
allows the use of multiple reference determinants. By considering the 
secondary references as excitations of the primary reference and using them 
to change the acceptance criteria for selection and spawning, we obtain a 
simple form of stochastic multireference coupled cluster which preserves the
 appealing aspects of the single reference approach. The method is able to 
successfully describe strongly correlated molecular systems using few 
references and low cluster truncation levels, showing promise as a tool to 
tackle strong correlation in more general systems. Moreover, it allows 
simple and comprehensive control of the included references and excitors
thereof, and this flexibility can be taken advantage of to gain insight 
into some of the inner workings of established electronic structure methods.

\end{abstract}
\maketitle

\section{Introduction}
The study of strong correlation in electron systems has been an important 
theme in electronic structure theory in recent years, as it is present in 
a series of interesting chemical systems, such as radicals, excited states,
transition states and dissociating bonds.\cite{Lyakh2011} 
In the presence of strong correlation, typically high-accuracy methods like 
coupled cluster often fail to correctly describe the system. This failure has
been attributed to the decrease in quality of the Hartree--Fock wavefunction 
as a first-order representation of the system, as the static correlation 
present often leads to near-degeneracies in the Hilbert space which cannot be 
captured by a single-determinant wavefunction. 
 
Coupled cluster (CC) theory\cite{Cizek1966,Cizek1969} has become the most 
popular \latin{ab initio} approach to electronic structure calculations, 
as it provides good results for medium-sized weakly correlated systems, 
while maintaining size-consistency and scaling polynomially with system size.
However, for strongly correlated systems, it 
requires consideration of high level excitors in order to correctly estimate
 the correlation energy.\cite{Chan2004} Since its computational 
costs scale as  $O(N^{2i+2})$, where $i$ is the truncation level and $N$ is 
the system size, this limits its use to very small systems.

One way to circumvent this issue and accurately treat some strongly correlated 
systems has been to use multiple reference determinants. Today, the field of
 multireference coupled cluster is very broad, with numerous methods 
developed over the last forty years, falling broadly into two categories:
particle-conserving methods, which either use multiple cluster operators
in a Jeziorski-Monkhorst \latin{ansatz}\cite{Jeziorski1981, 
Piecuch1992, Piecuch1994a, Mahapatra1998,Mahapatra1999, Kowalski2001a, Chattopadhyay2004, Hanrath2005}
or a single cluster operator,\cite{Mukherjee1975, Oliphant1991, Piecuch1993a, Piecuch1999, Kowalski2000,
Lyakh2005, Yanai2006, Fang2007, Li2008, Neuscamman2010, Evangelista2011, Shen2012} 
and Fock-space methods,\cite{Mukherjee1977, Haque1984, Lindgren1987, Stolarczyk1985,
Kaldor1991, Meissner1996, Meissner1998, Figgen2008} which generate 
wavefunctions with different numbers of electrons. 
While some of these have been successful in 
capturing the correlation energies of various test systems, \cite{Li2003, 
Balkov1991, Mukhopadhyay1991, Balkova1994,Das2010, Mukherjee1977} they are 
plagued by various size-consistency and intruder-state 
issues.\cite{Malrieu1985, Jankowski1994a, Kaldor1988, Kaldor1991, Paldus1993,
Piecuch1993, Piecuch1994, Kowalski2000, Kallay2002, Nooijen2005,Lyakh2008} 

In recent years, conventional quantum chemical techniques have been 
successfully combined with stochastic wavefunction propagation methods to 
improve computational performance. A prime example of this is the Full 
Configuration Interaction Quantum Monte Carlo (FCIQMC) 
method.\cite{Booth2009} While, like Full Configuration Interaction (FCI), 
FCIQMC scales exponentially with system size, it does so with a 
significantly lower prefactor. This has allowed the method, together with 
its initiator adaptation,\cite{Cleland2010} to successfully 
treat a variety of systems.\cite{Booth2011,Shepherd2012,Booth2013}

A stochastic solution to the coupled cluster equations has also been 
implemented using Projector Monte Carlo.\cite{Thom2010} This Coupled Cluster
 Monte Carlo (CCMC) method reproduces deterministic CC results to within 
stochastic error bars, but only needs to store a small fraction of the 
Hilbert space, leading to significantly lowered memory and computational 
costs. Both FCIQMC and CCMC have recently been used in conjunction with
conventional coupled cluster as a means to include selected higher-order 
clusters in a CC calculation, either iteratively or not.\cite{Deustua2017,
Deustua2018} 

In this paper we describe an implementation of multireference 
coupled cluster (using a single-reference formalism similar to that of 
\cite{Oliphant1991,Piecuch1993a}) within the stochastic paradigm,
which allows for very quick implementation of such methods.
In the following section we give an overview of the CCMC method and in the 
third section we describe our implementation of multireference Coupled 
Cluster Monte Carlo (mr-CCMC). The fourth and fifth sections then presents a series of 
results obtained using this method on known strongly correlated molecular 
systems. These results are discussed in the general context of multireference
methods in section 5 and some conclusions are given in section 6.

\section{Stochastic Coupled Cluster}
In deterministic CC, the wavefunction is represented by the exponential 
\latin{ansatz} 
\begin{equation}
\Psi_{\mathrm{CC}} = e^{\hat T}\ket{D_{0}},
\end{equation}
where $\ket{D_{0}}$ is 
a reference wavefunction (usually the Hartree--Fock wavefunction), 
\begin{equation}
\hat T = \sum_{\textbf i} t_{\textbf i}\hat a_{\textbf i}
\end{equation}
and $\hat a_{\textbf i}$ are excitors --- combinations of creation and annihilation 
operators (for example, a second order excited determinant $\ket{D_{ij}^{ab}} =
\hat a_b^\dagger \hat a_a^\dagger\hat a_i \hat a_j \ket{D_0}$, so $\hat a_{ij}^{ab} = 
\hat a_b^\dagger \hat a_a^\dagger\hat a_i \hat a_j$). If we group the excitors 
based on their excitation level relative to the reference, we can also write
\begin{equation}
\hat T = \sum_i \hat T_{i}
\end{equation}
where $i$ is the excitation level. This wavefunction is equivalent to the 
FCI wavefunction if all possible excitors are included. In truncated CC, 
$\hat T$ is limited to only excitors of up to a certain excitation level. 
In order to obtain $t_{\textbf i}$, the Schrödinger equation is projected 
onto each of the determinants $\ket{D_{\textbf i}}$ (including the reference),
leading to a series of coupled cluster equations to be solved:
\begin{equation}
\braket{D_{\textbf i}|\hat H - E|\Psi_{\mathrm{CC}}} = 0,
\end{equation} 
where $E$ is the energy of $\Psi_{\mathrm{CC}}$. The number and complexity 
of these equations increases with the highest excitation level considered.

These equations are equivalent to 
\begin{equation}
\braket{D_{\textbf i}|1-\delta\tau(
\hat H - E)|\Psi_{\mathrm{CC}}} = \braket{D_{\textbf i}|\Psi_{\mathrm{CC}}}
\end{equation}

Since 
\begin{equation}
\braket{D_{\textbf i}|\Psi_{\mathrm{CC}}} =\pm \braket{D_{0}|\hat 
a_{\textbf i}^{\dagger}|\Psi_{\mathrm{CC}}}=t_{\textbf i}+ O(\hat T^{2}),
\end{equation}
 this can be approximately recast in an iterative form as\cite{Spencer2016}
\begin{equation}
t_{\textbf i}(\tau + \delta\tau) = t_{\textbf i}(\tau) - \delta\tau\braket
{D_{\textbf i}|\hat H-E|\Psi_{\mathrm{CC}}}
\end{equation}

It is possible to obtain the solutions to these equations from the 
population dynamics of a set of `excips' in Hilbert space. 
This is done by stochastically sampling the action of the Hamiltonian,
described by two processes: spawning of an excip from 
$\ket{D_{\textbf i}}$ onto another  $\ket{D_{\textbf j}}$ coupled to it by 
the action of the Hamiltonian (with probability proportional to $\braket{D_{
\textbf j}|\hat H|D_{\textbf i}}$) and death of excips on $\ket{D_{\textbf i
}}$ (with probability proportional to $\braket{D_{\textbf i}|\hat H - 
S|D_{\textbf i}}$). The `shift' $S$  replaces the parameter $E$ in the 
stochastic coupled cluster equations. Finally, pairs of excips of opposite 
signs on the same excitor annihilate each other, which helps ensure that the
 algorithm converges on the correct nodal structure.\cite{Booth2009} In 
order to improve computational performance and stability, a series of 
modifications to this algorithm have been made, such as the deterministic 
selection of the reference and non-composite excitors,\cite{Spencer2018} the
implementation of an efficient importance-based selection 
method,\cite{Scott2017} the use of a similarity transformed Hamiltonian in the 
linked CCMC formalism\cite{Franklin2016} and the development of efficient 
excitation generators\cite{Holmes2016,Neufeld2018} and parallelizable algorithms.
\cite{Spencer2018} More recently a diagrammatic version of CCMC has been implemented.\cite{Scott2019}

From a CCMC calculation, we have two estimators for the correlation energy 
of $\ket{\Psi_{\mathrm{CCMC}}}$:\newline
\noindent 1. The instantaneous projected energy 
\begin{equation}E_{\mathrm{proj}} = \frac
{\braket{D_{0}|\hat H|\Psi_{\mathrm{CCMC}}}}{\braket{D_{0}|\Psi_{\mathrm{CCMC}}}}
\end{equation}
2. The `shift' $S$, which  is expected to converge to the correlation energy
 once the calculation has reached a stable excip population. Once a target 
population has been reached in a CCMC population, the shift is set to vary starting
from the instantaneous value of the projected energy. The shift is then updated every
$A$ steps using\cite{Booth2009}
\begin{equation}
S(\tau) =S(\tau - A\delta\tau) - \frac{\zeta}{A\delta\tau}\mathrm{ln}\frac{N_\mathrm{tot}(\tau)}{N_\mathrm{tot}(\tau-A\delta\tau)}
\end{equation}
where $N_\mathrm{tot}$ is the total excip population.

\section{Multireference Coupled Cluster}

Multireference methods are justified by the desire to include ``important" 
highly-excited determinants in the wavefunction expansion (e.g. configurations
with many electrons in antibonding orbitals that acquire large coefficients during bond breaking).
These are only included in the single reference CC (sr-CC) algorithm at high 
truncation levels. Their inclusion causes a significant improvement in the 
energy estimate (see Figure \ref{fig:multi}), but also requires an increased
 computational cost. However, by considering such determinants as part of the
 reference (or model) space of the calculation, they can be included without increasing 
the truncation level.
\subsection{Conventional MRCC}

Most Hilbert-space multireference coupled cluster methods are based on the 
Jeziorski-Monkhorst formalism\cite{Jeziorski1981}
\begin{equation}
\Psi_\mathrm{\nu} = \sum_\mu c_{\nu\mu} e^{\hat T^\mu}\Phi_\mu, 
\end{equation}
where $\Phi_\mu$ are the reference-space functions, $\hat T^\mu$ are cluster operators
defined relative to each of these references and $c_{\nu\mu}$ are CI coefficients. The formalism is the basis for so-called
\latin{state-universal} methods, where multiple wavefunctions are determined simultaneously.
\latin{State-specific} methods have also been developed; here we will discuss a particular
approach based on a single-reference formalism, known as SS CCSD(TQ)\cite{Oliphant1991, Oliphant1992, Piecuch1993a, 
Piecuch1994b} or CCSDtq.\cite{Piecuch1999} The starting point is to observe that for a given reference 
$\ket{D_\mathbf{i}} = \hat{a}_{i}\ket{D_0}$, 
\begin{equation}
e^{\hat T^{(\mathbf{i})}}\ket{D_\mathbf{i}} = e^{\hat T^{(\mathbf{i})}}\hat{a}_\mathbf{i}\ket{D_0}.
\end{equation}
It is therefore possible to rewrite any multireference wavefunction in terms of excitations of
a single reference only. With the appropriate intermediate normalisation ($\braket{D_0|\Psi} = 1$, 
it is further possible to write it in an exponential form, $e^{\hat T'}D_0$,
where as before $\hat T' = \sum_i \hat T_i'$, but highly excited $\hat T_i'$ no longer include all possible excitations
of order $i$ from $\ket{D_0}$, but only those that can be reached by lower excitations from other references. 
To clarify, let us consider a complete (2,2) reference space, given by the four determinants.
$\{ \ket{D_0}, \ket{D_I^A}, \ket{D_J^B}, \ket{D_{IJ}^{AB}}\}$. $I$ and $A$ may be taken to be $\alpha$ spin-orbitals, 
and $J$ and $B$ as $\beta$. If we are interested in including up to double excitations
out of this space in our cluster expansion, the wavefunction can be described as a single-reference exponential
\latin{ansatz}\cite{Piecuch1993a} with
\begin{equation}
\begin{split}
\hat T' &= \hat T_1 + \hat T_2 + \sum_{I',j,k,a,b,C'} \hat T_3\begin{pmatrix}abC' \\ I'jk\end{pmatrix} \\
&+\sum_{k,l,c,d,A,B} \hat T_4\begin{pmatrix}cdAB \\IJkl\end{pmatrix}
\label{ssccsdtq-t'}
\end{split}
\end{equation}
where model space orbitals have been labelled with capital
letters $I, J, K,...\ A, B, C...$ and general orbitals as $i, j, k...\ a, b,c...$. 
Primes have been used to differentiate summation indices over the model space
from fixed model space orbitals.
 
In general, the cluster operator $\hat T'$ can be written as a sum of internal
and external cluster operators,\cite{Piecuch1993a}
\begin{equation}
\hat T' = \hat T^\mathrm{int} + \hat T^\mathrm{ext}
\end{equation}
where $T^\mathrm{int}$ gives excitations within the model space and $T^\mathrm{ext}$
produces excitations outside the model space. 
\begin{align}
&\hat T^\mathrm{int} = \sum_{I',A'} \hat T_1\begin{pmatrix}A'\\I'\end{pmatrix} +
\hat T_2\begin{pmatrix}AB\\IJ\end{pmatrix} \\
&\hat T^\mathrm{ext} = \sum_{i,j,...,a,b,...} \hat T_1\begin{pmatrix}a\\i\end{pmatrix} +
\hat T_2\begin{pmatrix}ab\\ij\end{pmatrix} + ... 
\end{align}
where at least one of the indices in each cluster in $\hat T^\mathrm{ext}$ is not in the model space. 
The wavefunction may then be written as
\begin{equation}
\Psi = e^{\hat T^\mathrm{ext}}e^{\hat T^\mathrm{int}}\ket{D_0}
\end{equation}
It is worth pointing out at this stage that the singly excited terms in the model space are
constrained to have the correct spin (and therefore $\ket{D_I^B}$ and $\ket{D_J^A}$ are not included). However, 
pairs of model space orbitals of different spin can be included in 
$\hat T_3\begin{pmatrix}abC' \\ I'jk\end{pmatrix}$, so long as the overall spin of the terms is correct. 
In the case where all but two indices in the term come from outside the model space, such terms, 
while included in the \latin{ansatz}, are not related by a less than double excitation to any of
the model space determinants. This feature will be important 
when comparing to our stochastic method.

The \latin{ansatz} can be generalised to include any excitation level $\lambda$ from a model space of all at most
$k$-tuple excitations of a reference determinant, using a cluster operator\cite{Piecuch1993a}
\begin{equation}
\begin{split}
\hat T'& = \hat T_1 + ... + \hat T_\lambda + 
\hat T_{\lambda+1}\begin{pmatrix}a_1...a_\lambda A_1 \\ I_1 i_1...i_\lambda
\end{pmatrix} +\\
&+ ... +T_{\lambda + k} \begin{pmatrix}a_1...a_\lambda A_1...A_k \\ I_1...I_k i_1...i_\lambda
\end{pmatrix},
\end{split}
\label{eq:general}
\end{equation}

where indices $a_1,...,a_\lambda, i_1,...,i_\lambda$ correspond to active or inactive orbitals, while 
$A_1,...,A_k$, $I_1,...I_k$ are active indices.
\clearpage
\subsection{Stochastic MRCC}

Consider a stochastic coupled cluster calculation with truncation 
level $m$. Currently, the single reference algorithm selects clusters that 
correspond to an excitation of up to order $m+2$ of the reference and 
allows spawning onto those that correspond to excitations up to order $m$. 
We introduce a secondary reference in this model by allowing spawning and 
selection to occur in an expanded space, of size and shape determined by this secondary 
reference. However, the clusters will still be described exclusively by 
their effect on the primary reference, so we will not need to consider 
propagation differently for the two references. For example, in a system 
where the secondary reference considered is an excitation of order $n$ of the 
primary reference, we allow clusters to be selected if they correspond to 
excitations up to order $n+m+2$ of the primary reference. For high 
separations between references, this requires sampling a significantly 
larger space than the single reference equivalent, but due to recent 
improvements to the selection algorithm,\cite{Scott2017} this can be done 
relatively efficiently. Spawning is then only allowed onto excitors within 
$m$ excitations of \textit{either} of the references.

\begin{figure}[h]
\subfloat {\includegraphics[width=0.3\textwidth, center]{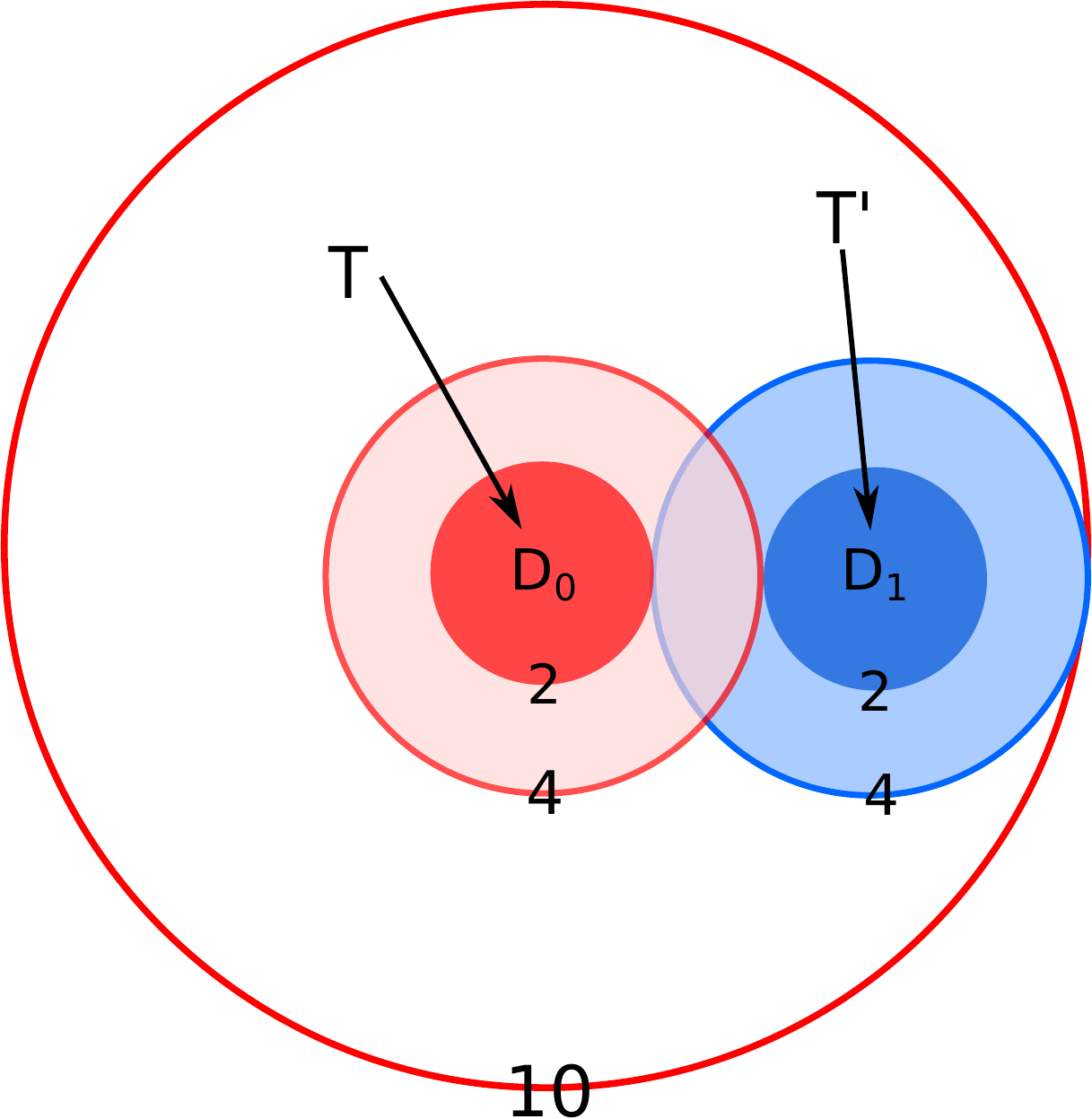}}
\caption {\protect\raggedright \footnotesize{Example of the explored space of a 
multireference stochastic CCSD calculation with the two references 6 
excitations apart. Clusters are selected within 10 excitations of 
$\ket{D_{0}}$ (inside the black circle), spawning attempts are made from clusters within 4 excitations
 of $\ket{D_{0}}$ or $\ket{D_{1}}$ (inside the transparent red/blue circles) to clusters within 2 excitations of $\ket
{D_{0}}$ or $\ket{D_{1}}$ (inside the solid red/blue circles). The regular stochastic CCSD calculation would only explore
the space inside the transparent red circle.}}
\label{fig:space}
\end{figure}

Figure \ref{fig:space} shows an example of two references 6 excitations apart,
treated at CCSD level. While this model nicely highlights the relation between
the selection and spawning spaces, for easier comparison to the conventional 
method of Piecuch, Oliphant and Adamowicz,\cite{Oliphant1991, Piecuch1993a} 
we will consider two references two excitations apart, $\ket{D_0}$ and $\ket{D_1} = \ket{D_{IJ}^{AB}}$
 and treat both at the CCSD level. The resulting wavefunction is given by $\Psi = e^{\hat{T'}} \ket{D_0}$,
where
\begin{@twocolumnfalse}
\begin{equation}
\begin{split}
\hat T' &= \hat T_1 + \hat T_2 
+ \sum_{A',I', a, b, j, k, l} \Big[\hat T_3\begin{pmatrix}aAB\\IJk\end{pmatrix} \\ 
&+ \hat T_3\begin{pmatrix}abA'\\IJk\end{pmatrix} + \hat T_3\begin{pmatrix}aAB\\I'jk\end{pmatrix}
+ \hat T_4\begin{pmatrix}abAB\\IJkl\end{pmatrix} \Big]
\end{split}
\label{eq:2r-t'}
\end{equation}
\end{@twocolumnfalse}
where $A'$ runs over model orbitals empty in $\ket{D_0}$, $I'$ runs over occupied ones, $j,k,l$ are core orbitals and $a,b$ are virtual orbitals.
In terms of internal and external cluster operators,
\begin{equation}
\hat T^\mathrm{int} = \hat T_2\begin{pmatrix}AB\\IJ\end{pmatrix}
\end{equation}
and $\hat T^\mathrm{ext} = \hat T' - \hat T^\mathrm{int}$.
Obviously, 
\begin{equation}
\hat T_2\begin{pmatrix}AB\\IJ\end{pmatrix} \propto \hat{a}_1,
\end{equation}
where $\hat{a}_1\ket{D_0} = \ket{D_1}$. Also if we label $\hat T^{(1)}$ cluster operators relative to $\ket{D_1}$, then
\begin{align}
&\hat T_3\begin{pmatrix}aAB\\IJk\end{pmatrix} \propto \hat T_1^{(1)}\begin{pmatrix}a\\k\end{pmatrix}\hat{a}_1\\
&\hat T_3\begin{pmatrix}abA\\IJk\end{pmatrix} \propto \hat T_2^{(1)}\begin{pmatrix}ab\\Bk\end{pmatrix}\hat{a}_1
\end{align}
and so on. We can therefore write
\begin{equation}
\hat T' = \hat T_1 + \hat T_2 + (\hat T_1^{(1)} + \hat T_2^{(1)})\hat{a}_1
\end{equation}
taking care to only include overlapping contributions in the cluster operator relative
 to a single reference. This leads to multiple equivalent representation of $\hat T'$
in this form. In our stochastic approach, we only consider each excited determinant 
once, regardless of which references and clusters it can be reached from, so these considerations are 
trivially avoided. This makes CCMC an ideal framework for this kind of algorithm, as it 
allows simple implementations of potentially complicated reference spaces.
Generalizing to an arbitrary number of references, with arbitrary corresponding truncation levels,
we obtain
\begin{equation}
\hat T'=\sum_{i=1}^{m_0}
\hat T_{i} + \sum_{n=1}^{N}\sum_{j=0}^{m_n}\hat T^{(n)}_{j}\hat a_n
\end{equation}
where $\hat T_i$ are $i$-th order excitors of the first reference, $\hat T^{(n)}_j$ 
are $j$-th order excitors of the $n$-th secondary reference, $\hat a_n$ is the 
excitor that generates the $n$-th secondary reference from the first, $m_n$ is the
truncation level for reference $n$ and $N$ is the number of secondary references used. 
We note here two differences from Eq. \ref{eq:general}. Firstly, our formalism allows
the definition of an arbitrary reference space, rather than requiring the inclusion of all
excitations up to a certain order. Secondly, the truncation level with respect to each reference
can be selected independently, allowing for additional flexibility.

This algorithm effectively allows consideration of secondary references while 
maintaining the relative simplicity of the sr-CCMC approach.  It is worth 
noting that, in a multireference calculation that explores the set of 
determinants within $m$ excitations of two references, there is an 
approximately twofold increase in the proportion of the Hilbert space that 
must be stored compared to the corresponding single-reference calculation, 
truncated at excitation level $m$. In general, we expect the space spanned by
a calculation to increase at most linearly with the number of references, provided
the truncation levels are the same for all references, as the spaces spanned
by the cluster expansion about each reference may overlap, leading to slight sublinearity.
In large basis sets, this is insignificant relative to the $\mathcal O(N^{2n})$ increase in memory costs associated with 
increasing the truncation level to $m+n$ in order to include the same 
determinants in a single-reference calculation. If lower truncation levels 
can be used to obtain results of the same accuracy, the scaling with system 
size is reduced polynomially. With the current selection scheme, the size of 
the selection space is only determined by the highest excited secondary reference, so we expect 
the computational scaling with number of references to be favourable.

If we include the two single excitations $\ket{D_I^A}$ and $\ket{D_J^B}$ in the model space, 
equation \eqref{eq:2r-t'} becomes
\begin{equation}
\begin{split}
\hat T' &= \hat T_1 + \hat T_2 
+ \sum_{A',I', a, b, j, k, l} \Big[\hat T_3\begin{pmatrix}aAB\\IJk\end{pmatrix} \\
&+ \hat T_3\begin{pmatrix}abB\\Jkl\end{pmatrix}
+ \hat T_3\begin{pmatrix}abA\\Ijk\end{pmatrix} \\
&+ \hat T_3\begin{pmatrix}abA'\\IJk\end{pmatrix} + \hat T_3\begin{pmatrix}aAB\\I'jk\end{pmatrix}
+ \hat T_4\begin{pmatrix}abAB\\IJkl\end{pmatrix} \Big]
\end{split}
\label{eq:4r-t'}
\end{equation}
If we compare this to equation \eqref{ssccsdtq-t'}, we find all terms are accounted for, except 
for those of the form $\hat T_3\begin{pmatrix}abB\\Ikl\end{pmatrix}$ and 
$\hat T_3\begin{pmatrix}abA\\Jkl\end{pmatrix}$. This is because, as mentioned as the end of 
the previous section, these are not within two excitations of any of the references. We therefore 
expect that, depending upon the magnitude of the contributions of such terms to the wavefunction, 
we may be able to observe differences between the SS CCSD(TQ) method and the mr-CCMCSD method, 
even when using the same model space. It is also worth pointing out that, while in our case there
 is a difference between equations \eqref{eq:2r-t'} and \eqref{eq:4r-t'}, both of these model spaces
would be described by the same cluster operator in SS CCSD(TQ), as the set of active orbitals is unchanged.
 Only the formal split of excitors between $\hat T^\mathrm{int}$ and $\hat T^\mathrm{ext}$ would change.

\section{Two-Reference Results}
\subsection{The S$_4$ model}

First we look at a simple 4-electron system, known as the S$_4$ model --- $\text{H}_4$ 
in a square geometry,\cite{Paldus1993}. The symmetry of the 
system and the fact that each of the H-H distances may be longer than an 
equilibrium $\text{H}_2$ bond introduces significant electron 
correlation to the system, so we expect it to have some multireference 
character. As we increase the H--H separation, while maintaining the square geometry, the
amount of strong correlation in the system also increases.

In a minimal basis,\cite{Jankowski1980} this system only has 10 Slater 
determinants in its Hilbert space, so we can easily obtain the FCI energy. 
In this case, both CCSDTQ and mr-CCSD with two references (2r-CCMCSD), 
where the second reference is chosen to be the highest excited determinant,
explore the entire Hilbert space, so we expect very good agreement of 
both methods with FCI. Therefore this system is a good test that the behaviour
of our algorithm is as expected. We can see from Table \ref{tab:H} that, at 
$r_\mathrm{HH} = 2 a_{0}$ there is
indeed good agreement between the FCI result, CCMCSDTQ and 2r-CCMCSD projected energies, 
with differences of less than 0.1 milliHartrees, well within chemical 
accuracy ($1.6\times10^{-3}$ Hartree). Our results compare favourably to conventional MRCC results obtained for
this system\cite{Balkov1991,Paldus1993,Piecuch1993}.
\begin{table}[h]
\footnotesize
\begin{center}
\begin{tabular}{|c|c|}
\hline
\textbf{Method}&\textbf{Energy/$\mathrm{E}_{h}$}\\
\hline
FCI & -0.117621\\
CCMCSD & -0.12044(3)\\
CCMCSDT & -0.12059(7)\\
CCMCSDTQ & -0.11761(7)\\
2r-CCMCSD & -0.11763(4)\\
MRCCSD\cite{Balkov1991} & -0.117580\\
MRCCSD-1\cite{Paldus1993} & -0.117686\\
MRCCSD-2,3\cite{Paldus1993} & -0.117575\\
MRACPQ-1\cite{Piecuch1993} & -0.117102\\
\hline
\end{tabular}
\end{center}
\caption{\protect\raggedright \footnotesize Values of the calculated correlation 
energy for H$_{4}$ in a minimal basis.}
\label{tab:H}
\end{table}

\begin{figure}[h]
\subfloat {\includegraphics[width=0.5\textwidth, center]{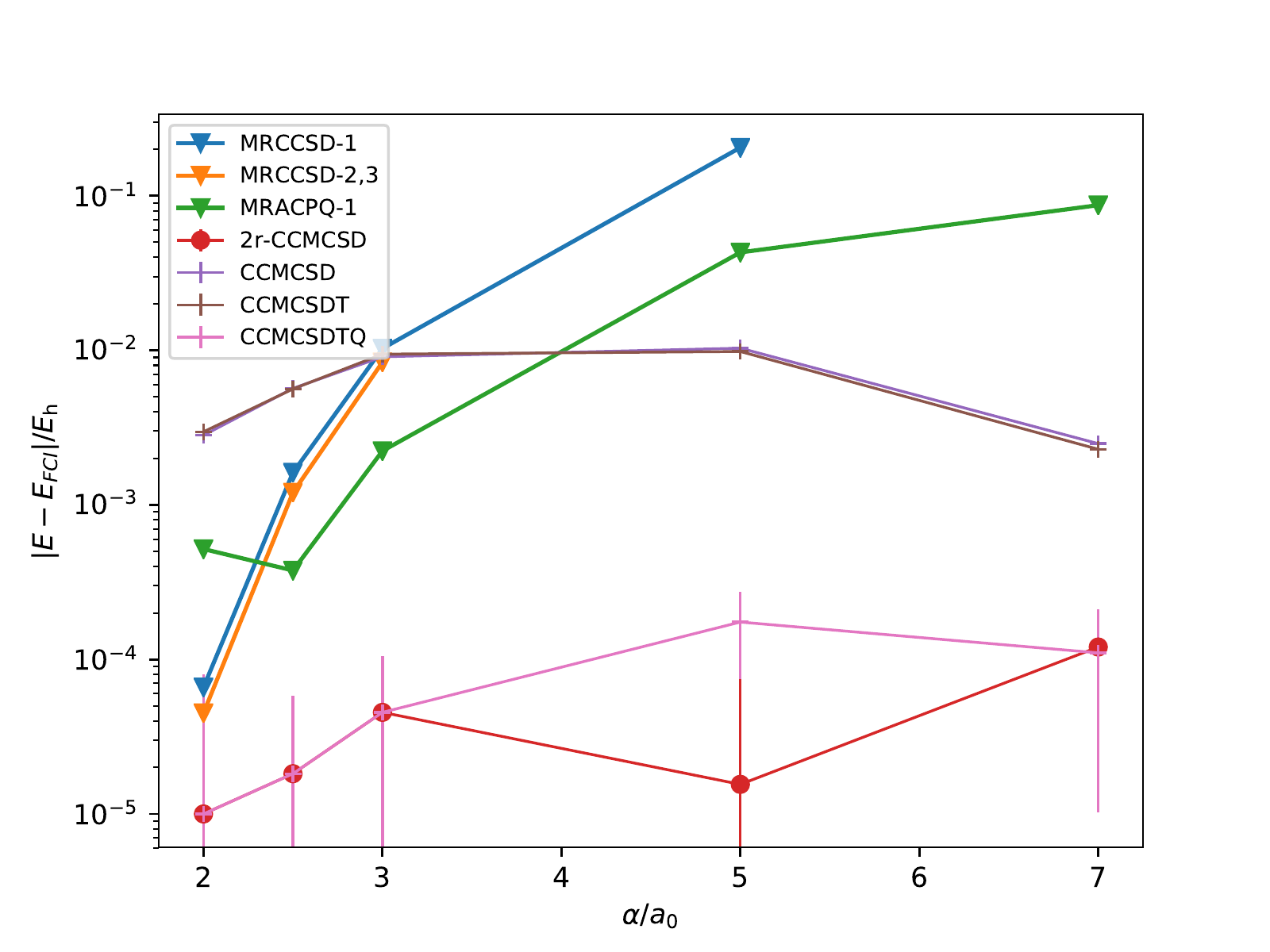}}
\caption {\protect\raggedright \footnotesize{Difference in correlation energy captured by various single 
and multireference coupled cluster methods relative to FCI, as we increase the length of the side of the H$_4$ square, $a$. As expected CCSDTQ
and 2r-CCSD are both in very good agreement with FCI over the full range of geometries investigated.}}
\label{fig:S4}
\end{figure}

A range of conventional MRCC methods\cite{Paldus1993,Piecuch1993} have been used
to investigate this system in its strongly correlated regime ($\alpha/a_0 \in [2,7]$).
As can be seen in Figure \ref{fig:S4} the quality of our energy estimate remains 
consistent over this interval, showing an order of magnitude improvement over previous results.

\subsection{The N$_2$ molecule}
\begin{figure}[h!]
\subfloat {\includegraphics[width=0.4\textwidth]{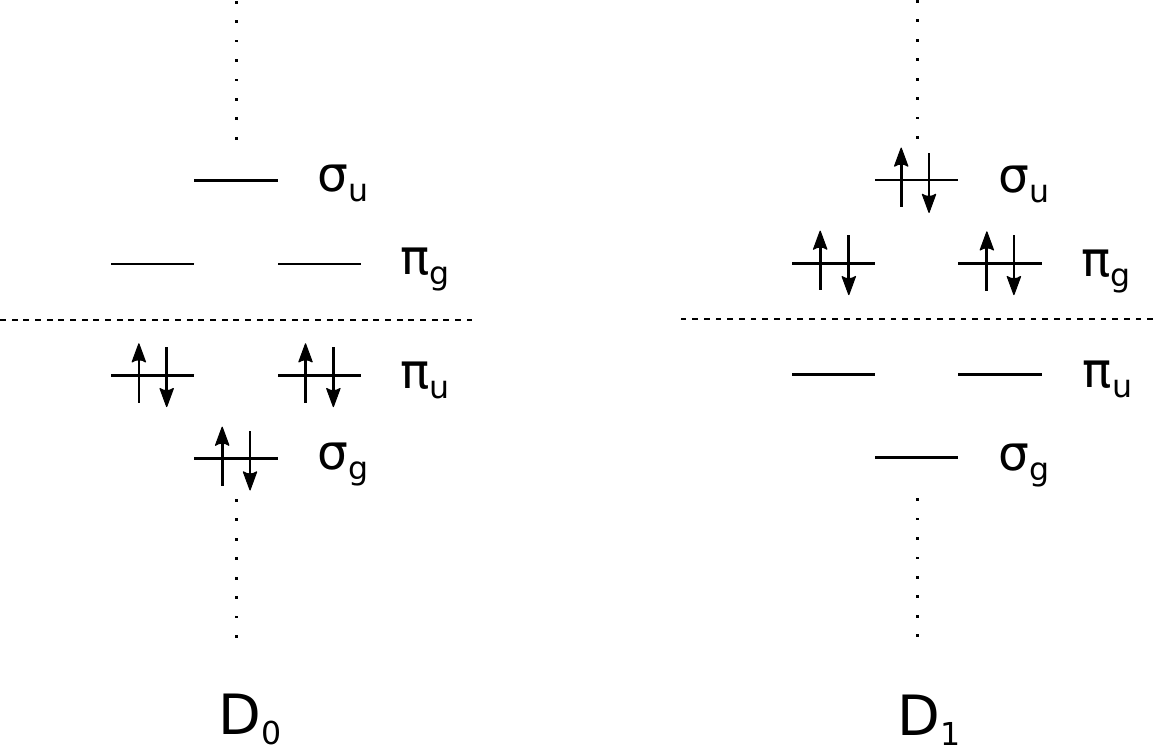}}
\caption{\protect\raggedright \footnotesize Active orbitals occupied in the
two references used for the nitrogen molecules.}
\label{fig:MO}
\end{figure}

\begin{figure}[h!]
\centering
\subfloat {\includegraphics[width=0.5\textwidth]{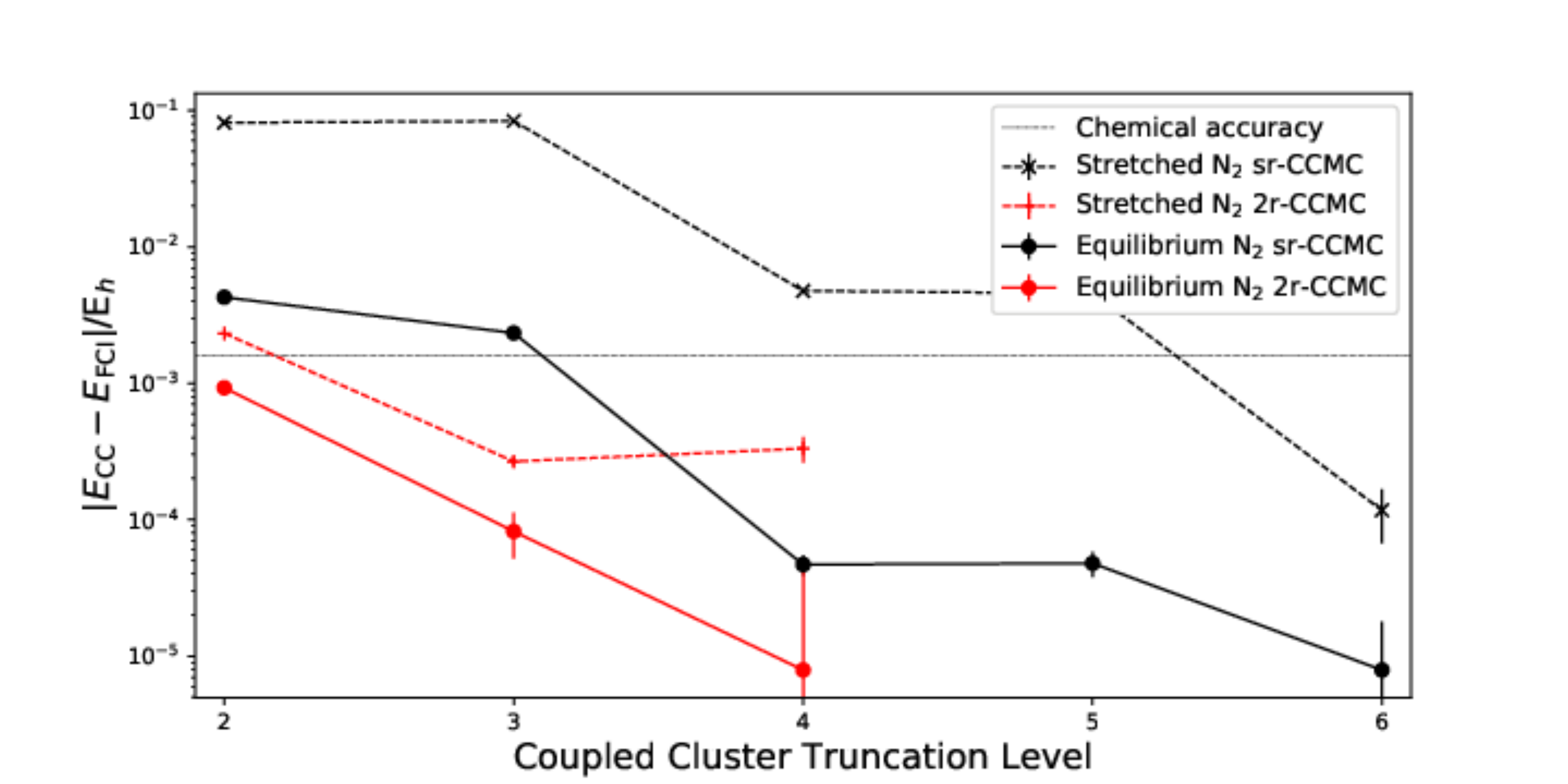}}
\\
\subfloat {\includegraphics[width=0.5\textwidth]{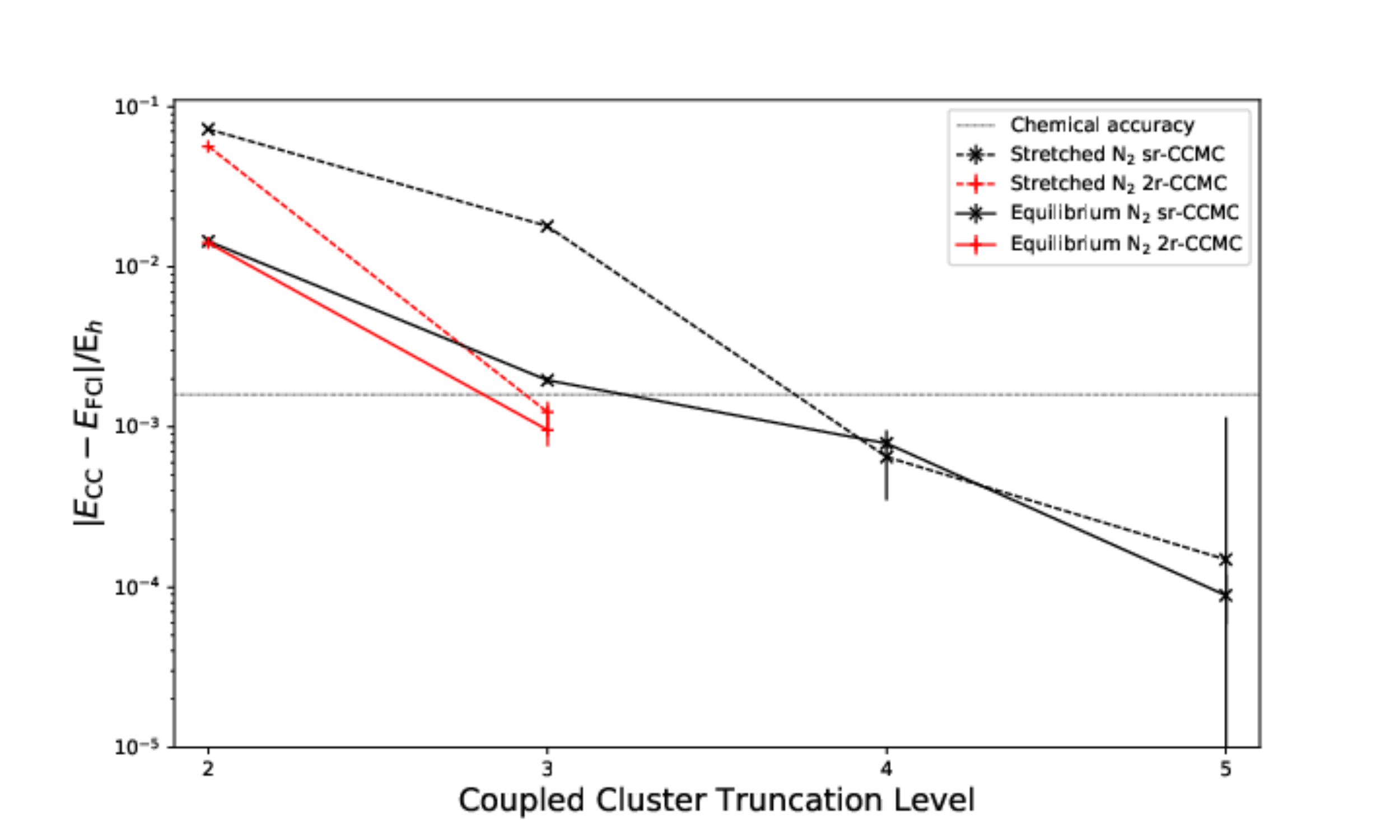}}
\caption{\protect\raggedright \footnotesize The difference between the coupled 
cluster and the FCI energy for different CC truncation levels for N$_{2}$ in
 a minimal basis (top) and in a Dunning cc-pVDZ basis(bottom) with frozen 
core electrons. For the larger basis set, the shift was used as a 
correlation energy estimator rather than the projected energy, due to 
difficulties collecting statistics on the latter. While in the 
single-reference case high truncation levels are needed to obtain 
sub-milliHartree accuracy, 2r-CCMCSDT is sufficient to achieve this.}
\label{fig:multi}
\end{figure}

The next system of interest is $\text{N}_{2}$, which is known to be 
difficult to accurately describe by single reference methods at stretched 
geometries, due to correlation effects caused by the dissociation of the 
triple bond.\cite{Laidig1987, Chan2004} Going from the equilibrium bond 
length ($2.118 a_{0}$) to $3.6 a_{0}$, the convergence of the coupled 
cluster energy with truncation level becomes significantly poorer (Figure 
\ref{fig:multi}), requiring costly, high-truncation level calculations to 
converge on the FCI result. mr-CCMC can be applied to this 
system, using a sixth order excitation of the Hartree--Fock determinant as 
our second reference. This corresponds to exciting six electrons from 
bonding $\sigma$ and $\pi$ orbitals to anti-bonding ones (see Figure \ref{fig:MO}). We expect that 
this determinant is crucial in describing the bond breaking that occurs as 
the nitrogen molecule is stretched and therefore a good candidate for a 
second reference.


The numerical results of single- and multireference calculations on 
stretched nitrogen are given in the Supporting Information. For reference,
Hartree--Fock energies are also given. In a STO-3G basis,\cite{Hehre1969} 2r-CCMC provides
 a significant improvement to our energy estimates, making 2r-CCMCSDT sufficient 
to get within chemical accuracy of the 
calculated FCI energy (Figure \ref{fig:multi}). A similar improvement can 
also be observed when treating the molecule in a larger Dunning cc-pVDZ 
basis set\cite{Dunning1989} with frozen core electrons (Figure 
\ref{fig:multi}), confirming that the faster convergence is not simply a 
consequence of the multireference space effectively covering a high 
proportion of the relatively small STO-3G Hilbert space. 

\begin{figure}[h]
\subfloat {\includegraphics[width=0.5\textwidth, center]{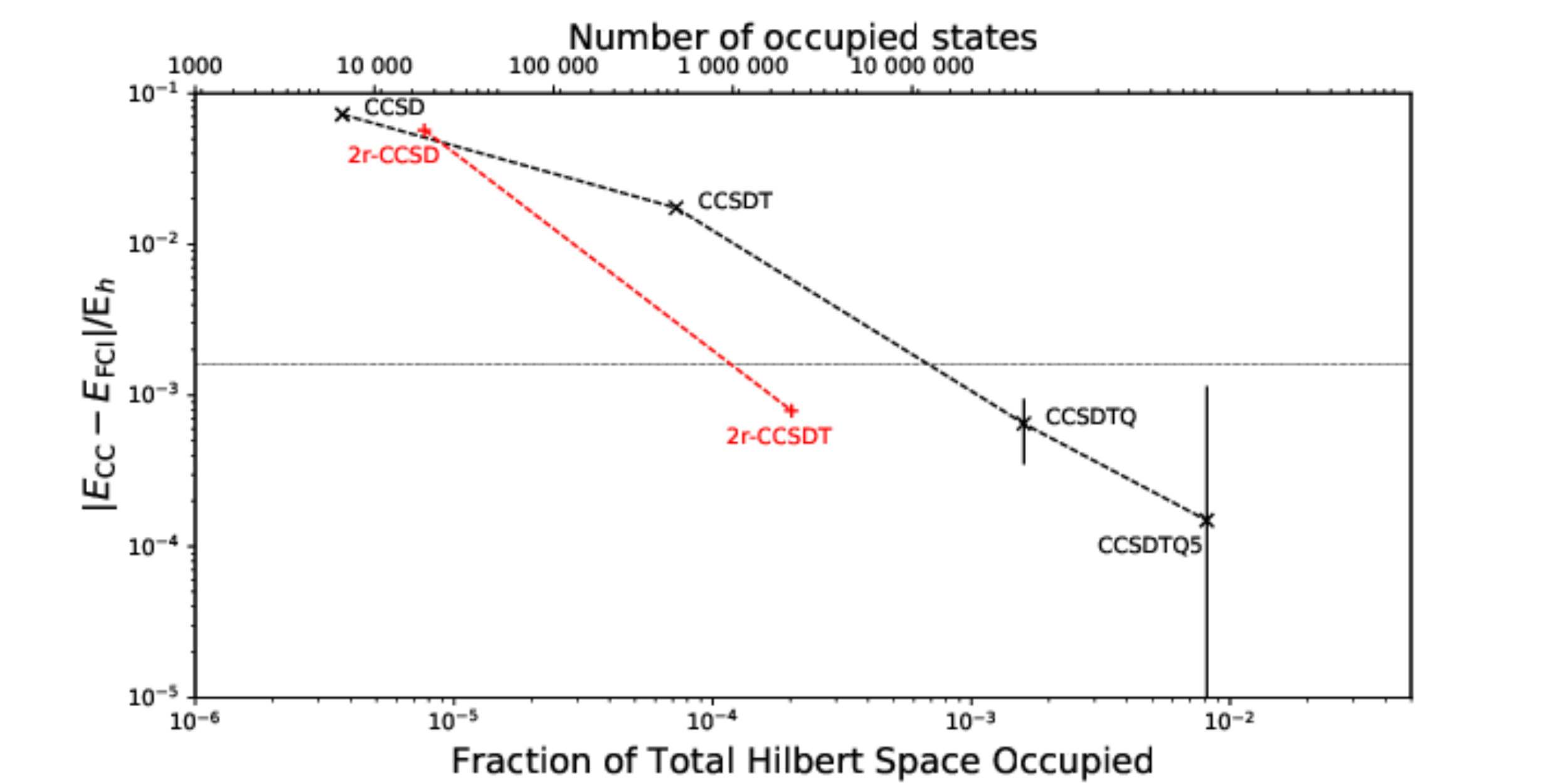}}
\caption {\protect\raggedright \footnotesize Convergence of CCMC energy versus the 
proportion of the total Hilbert Space that is populated once steady-state 
has been reached in Dunning cc-pVDZ stretched N$_{2}$, with frozen core 
electrons. The points correspond to successive truncation levels, starting 
at CCSD. It can be seen that 2r-CCMC achieves higher accuracy results with 
only a fraction of the memory requirements of high-level single-reference 
calculations.}
\label{fig:mem}
\end{figure}

Figure \ref{fig:mem} shows the proportion of the Hilbert space populated 
after the system has reached steady-state for single and multireference 
calculations. In general, coupled cluster memory costs should scale as
$\mathcal O(N^{2l})$ and even calculations with high truncation levels only
use a small fraction of the full Hilbert space of the system.
 Stochastic methods decrease the memory cost by a
constant pre-factor \cite{Spencer2016}.
It can be seen from Figure \ref{fig:mem} that 2r-CCMC produces more accurate 
results at a reduced memory cost relative to single-reference CCMC. 
Also, if CCSDT can be used to obtain results of similar accuracy to CCSDTQ, 
this reduces the scaling with 
system size by a factor of $N^{2}$ ($N^{6}$ vs. $N^{8}$),
provided an efficient sampling method for the multireference space is implemented.

\begin{figure}[h]
\subfloat {\includegraphics[width=0.5\textwidth, center]{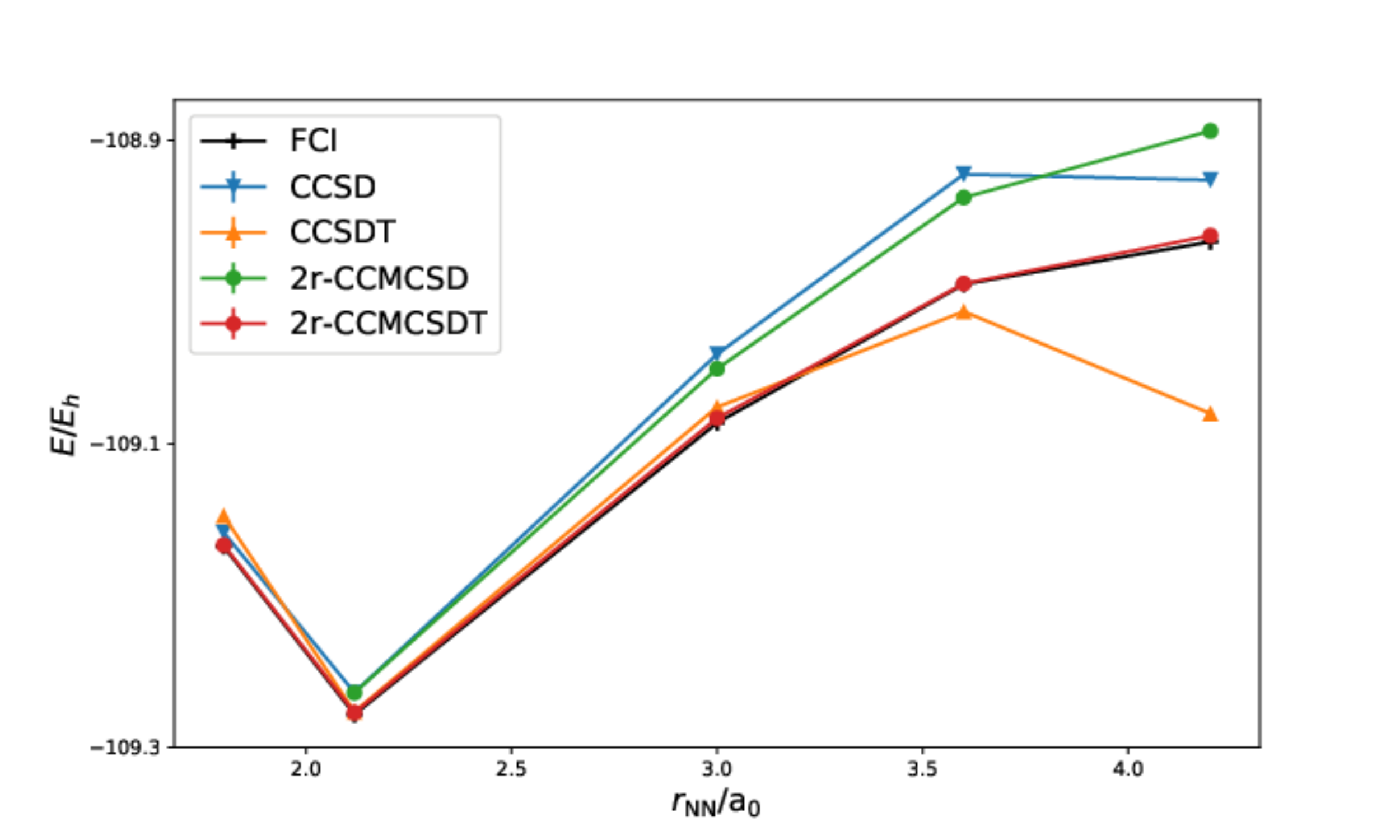}}
\caption {\protect\raggedright \footnotesize N$_2$ binding curves obtained using the
 Dunning cc-pVDZ basis set. At large separations ($r = 4.2 a_0$), the 
two-reference solution is metastable, with a long enough lifetime to 
collect statistics. Single-reference CCMC calculations are shown aside from 
$r = 4.2 a_0$ where deterministic CC values from Ref. \citenum{Chan2004} are given 
as the CCMC calculations are unstable, and $r = 1.8 a_0$  where a 
deterministic CC calculation was performed. FCI results are from 
Ref. \citenum{Chan2004}, except for at $r = 1.8 a_0$, where an  FCIQMC calculation 
was performed.}
\label{fig:binding}
\end{figure}

We have also used mr-CCMC to calculate a binding curve for N$_2$, given in 
Figure \ref{fig:binding}. Curves obtained from 2r-CCMCSDT are in 
significantly better agreement with FCI values\cite{Chan2004} than CCSD or
CCSDT. These results will be discussed further in the following section. 

\subsection{The N$_3^-$ anion}
\begin{figure}[h]
\centering
\subfloat {\includegraphics[width=0.5\textwidth]{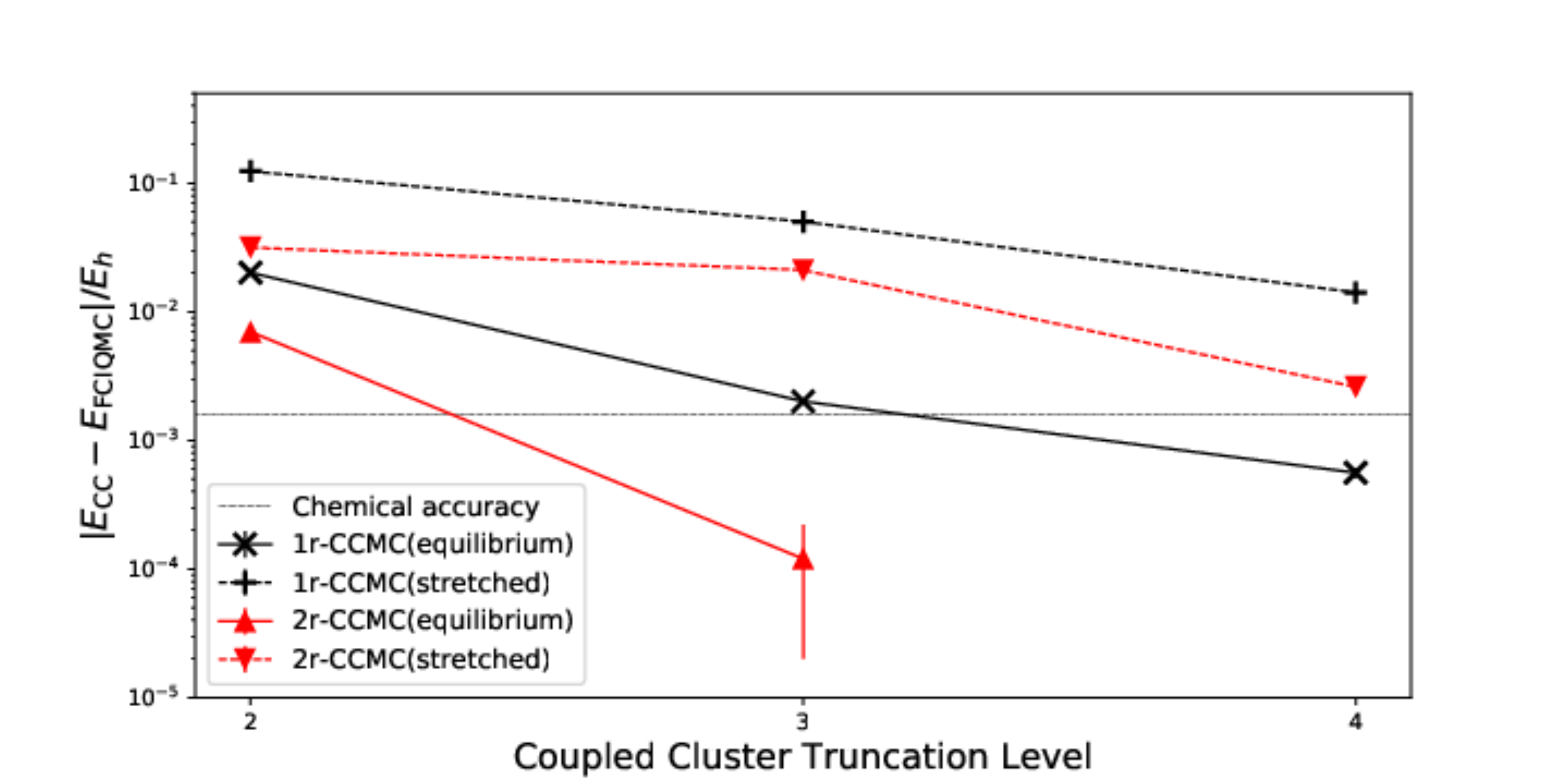}}
\caption{\protect\raggedright \footnotesize The difference between the CCMC 
and the FCIQMC energy for different CC truncation levels for 
N$_{3}^{-}$ in a minimal basis, at equilibrium and stretched geometries. 
Multireference CCMC provides a systematic improvement of the energy in both 
cases.}
\label{fig:n3}
\end{figure}

Finally, we look at the azide anion in order to assess the effect of using a
 second reference in systems with larger numbers of electrons. We have found
 that both the equilibrium geometry 
($r_{\mathrm{NN}}=1.16$ \r{A})\cite{jones2016chemistry} and a linear
symmetrically stretched geometry 
($r_{\mathrm{NN}}=2.0$ \r{A}) require high truncation levels for the 
CCMC energy to converge. For the multireference calculations, a quadruple
 excitation was used as the second reference, corresponding to the excitation 
of the four electrons in the non-bonding $\pi$ orbitals to the corresponding
antibonding orbitals. As can be seen in Figure \ref{fig:n3}, once 
again 2r-CCMC provides a significant improvement to the energy estimate, 
even if 2r-CCMCSDT is not sufficient to reach chemical accuracy in the stretched 
case.

The poorer convergence for 2r-CCMC for N$_3^-$ suggests that the choice 
of secondary reference has a significant effect on the quality of the 
results. This is as expected, following from the notion that references 
should be highly weighted determinants in the expansion of the true ground 
state wavefunction. In the case of N$_{2}$ we were aware of such a 
determinant, but for N$_3^{-}$, we have at multiple reasonable choices of 
secondary reference, one of which is the fourth order excitation used. 
However, given that this excitation is already significant in 
the equilibrium geometry, it is likely that upon stretching the bonds, more 
highly excited determinants (perhaps the one corresponding to the 
excitation of both $\sigma$ and $\pi$ electrons, as for N$_2$, or the excitation
of bonding rather than non-bonding $\pi$ electrons)
become highly weighted in the ground state and would therefore 
serve as better secondary references. 

\section{Beyond two references}
We have shown in the previous section that, using two references, mr-CCMC is 
more successful in capturing the correlation in difficult molecular systems 
than the corresponding single reference methods. In this section we will 
turn our attention to the performance of the method relative to conventional 
MRCC methods and propose a procedure to balance the accuracy of our method 
against its computational cost.

\subsection{Comparison to conventional MRCC methods}
\subsubsection{The H$_8$ model}
\begin{figure}[h]
\subfloat {\includegraphics[width=0.3\textwidth, center]{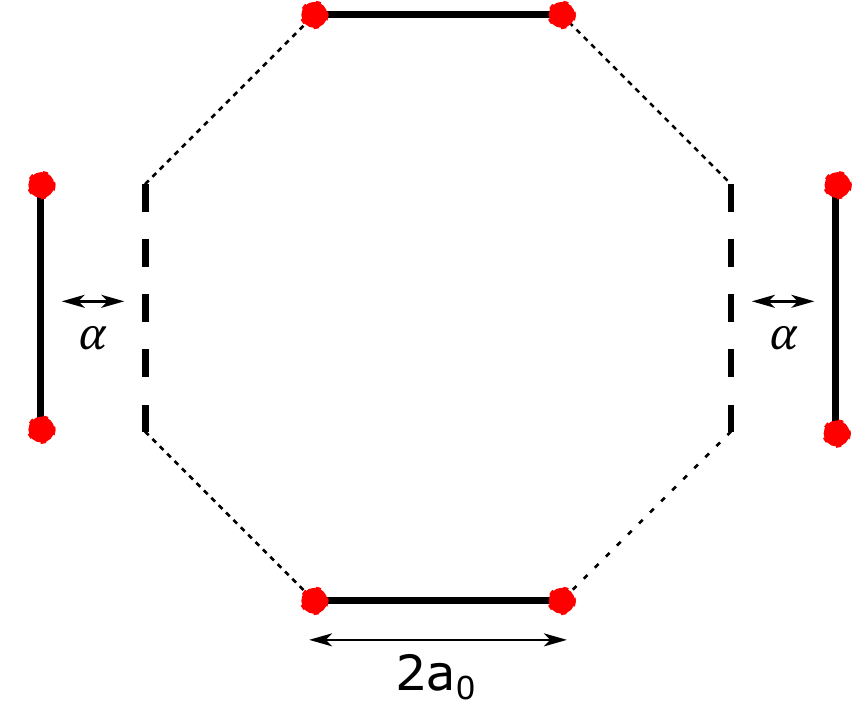}}\\
\caption {\protect\raggedright \footnotesize Geometry of the H$_8$ model.}
\label{fig:octo}
\end{figure}

\begin{figure}[h]
\subfloat {\includegraphics[width=0.5\textwidth, center]{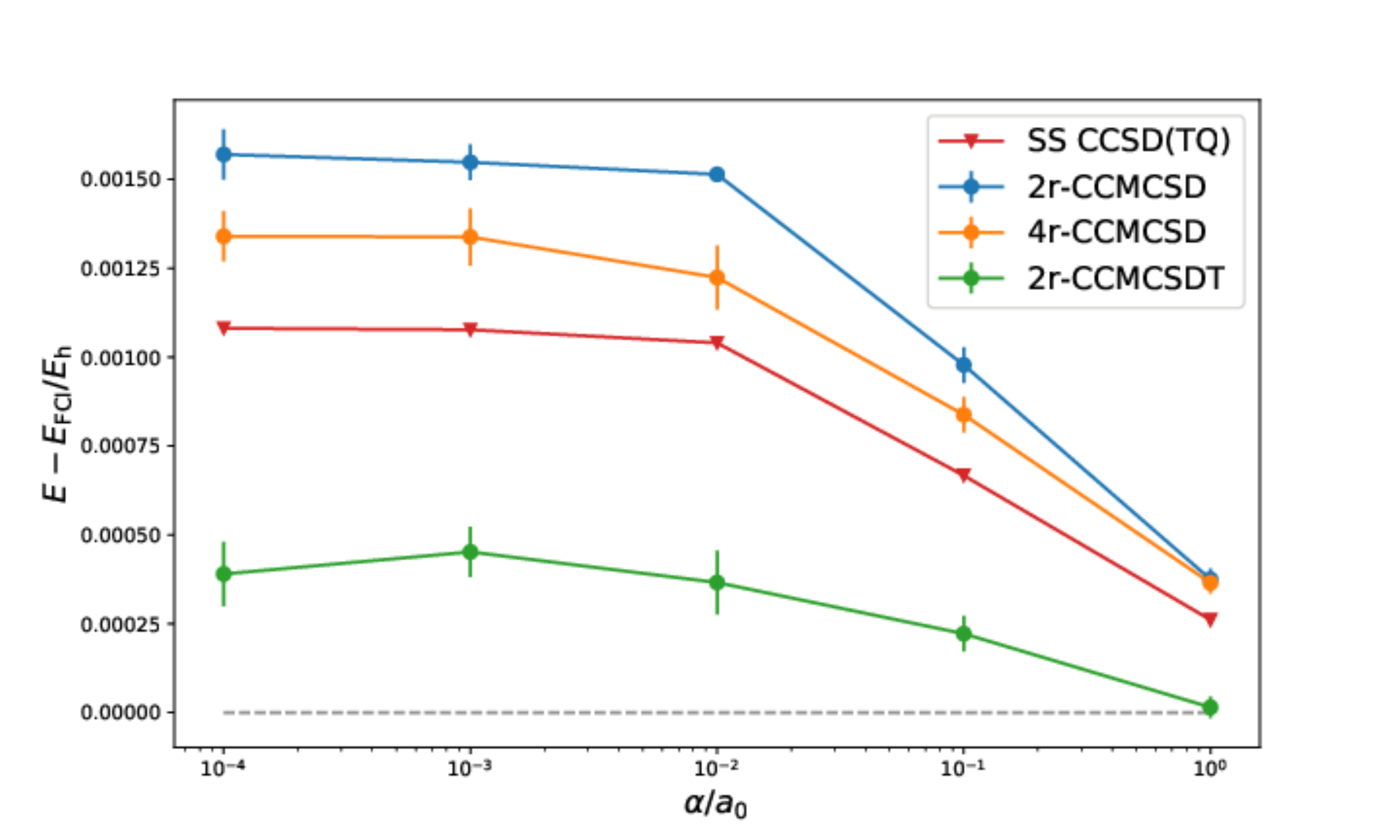}}\\
\caption {\protect\raggedright \footnotesize Comparison of single- and 
multireference coupled cluster results for the H$_8$ model.}
\label{fig:H8}
\end{figure}

First, we turn our attention to the H$_8$ model,\cite{Jankowski1985} shown in Figure \ref{fig:octo}, 
in a minimal basis.\cite{Jankowski1980} 
As the parameter $\alpha$ is varied from 0 to $\infty$, the degree of 
electron correlation in the system decreases, as it dissociates to 4 
independent H$_2$ molecules. The HOMO and LUMO of the system at 
$\alpha=a_0$ become closer in energy as we decrease $\alpha$, tending 
to degeneracy at $\alpha = 0$. Therefore, these orbitals form a natural 
choice of model space for mr-CCMC. This system has been studied using 
the SS CCSD(TQ) method of Piecuch, Oliphant and Adamowicz,\cite{Piecuch1994b} 
allowing a direct comparison. We investigate the system for 
$\alpha/a_0 \in [0.0001,1]$ and the results are given in Figure \ref{fig:H8}.
We find a noticeable ($\approx 0.5$ miliHArtree) discrepancy between 2r-CCMC 
and SS CCSD(TQ) energies. Including the full(2,2) CAS in the reference space 
decreases this discrepancy, but 4r-CCMCSD is still not in agreement with
SS CCSD(TQ). We expect this to be due to the absence of some terms in our 
cluster expansions compared to SS CCSD(TQ). To verify this, we compare the 
values given by these different approaches for the leading $\hat T_3$ terms 
in the CCSDTQ expansion.

\begin{center}
\begin{table*}[t]
\footnotesize
\begin{threeparttable}
\begin{tabular}{|c|c|c|c|c|c||c|}
\hline
\multirow{2}{*}{$\alpha/a_0$} & \multirow{2}{*}{$\hat T_3$ term} &
\multicolumn{5}{c|}{Coefficient$^\mathrm{a}$}\\ \cline{3-7}
& & CCSDTQ$^\mathrm{b}$ & SS-CCSDTQ$^\mathrm{b}$ & 2r-CCMCSD$^\mathrm{c}$ & 4r-CCMCSD$^\mathrm{c}$ & 2r-CCMCSDT$^\mathrm{c}$\\ \hline
\multirow{5}{*}{1.0} & $t_{875}^{653}$ & $0.0027990300$ & $0.0$ & $0.0$ & $0.0$ & $-0.006938820228$\\ 
 & $t_{873}^{543}$ & $-0.0027927466$ & $0.0$ & $0.0$ & $0.0$ & $-0.005822500732$\\ 
 & $t_{871}^{853}$ &  $0.0026117601$ & $0.0$ & $0.0$ & $0.0$ & $-0.00622210191$\\ 
 & $t_{875}^{741}$ & $-0.002605554$ & $-0.0026078696$ & $0.004204674226$ & $0.006681790253$ & $0.005893986503$\\ 
 & $t_{764}^{872}$ &  $0.0025749909$ & $0.0026218416$ & $0.0$ & $0.0$ & $-0.005829632611$\\ 
\hline
\multirow{6}{*}{0.1} & $t_{875}^{321}$ & $-0.0353278703$ & $-0.0304035827$ & $-0.02906899421$ & $-0.02533820583$ & $-0.03510182677$\\ 
 & $t_{873}^{521}$ & $0.0341886609$ & $0.0294198501$ & $-0.03320221354$ & $-0.2576256265$ & $-0.03417247948$\\ 
  & $t_{763}^{721}$ & $0.0112172942$ & $0.0091712186$ & $0.009037433273$ & $0.01719834927$ & $0.01032064989$\\ 
  & $t_{754}^{721}$ & $-0.0109948657$ & $-0.0089897429$ & $-0.009720349583$ & $-0.009384567264$ & $0.01176631082$\\ 
  & $t_{732}^{521}$ & $-0.0108251409$ & $-0.0087530913$ & $0.007848827095$ & $0.01093788654$ & $0.005768813771$\\ 
  & $t_{764}^{872}$ & -- & -- & $0.0$ & $0.0$ & $0.007187521769$\\ 
\hline
\end{tabular}
\caption{\protect\raggedright \footnotesize Leading triply excited cluster
 coefficients in the H$_8$ wavefunction for $\alpha/a_0 = 1.0,0.1,0.0001$.\\
$^\mathrm{a}$  All wavefunctions are normalised such that $\braket{D_0|\Psi} = 1$\\
$^\mathrm{b}$ As given in \cite{Piecuch1994b}\\
$^\mathrm{c}$ Values taken from instantaneous snapshots of the stochastic wavefunction.}
\label{tab:H8}
\end{threeparttable}
\end{table*}
\end{center}

As can be seen from Table \ref{tab:H8}, we observe more sign differences between
the mr-CCSD wavefunctions and CCSDTQ than when comparing to SS CCSD(TQ). Also,
at $\alpha = a_0$, the fifth largest triple excitation coefficient in CCSDTQ 
is on a term that is excluded from our calculations, but included in SS CCSD(TQ). 
The presence of such terms in the wavefunction at other geometries as well could
explain the discrepancy between our methods and SS CCSD(TQ). 
Indeed if we look at the 2r-CCSDT wavefunction (which includes some pentuple
contributions and therefore has significantly different cluster amplitudes than
CCSDTQ), this cluster continues to make a significant contribution at $\alpha = 0.1a_0$.
We can therefore expect such clusters to continue being significant in the CCSDTQ and SS CCSD(TQ)
wavefunctions, potentially justifying the discrepancy with relative to 2r- and 4r-CCMCSD.
While our method does not include them when considering the active space, 
a small number of additional references could be included to ensure their presence. 
Alternatively, increasing the truncation level to CCSDT in a two reference calculation is 
sufficient to recover these terms and indeed obtain much more accurate results than either 
mr-CCMCSD or SS CCSD(TQ).

\subsubsection{The N$_2$ molecule}
\begin{figure}[h]
\subfloat {\includegraphics[width=0.5\textwidth, center]{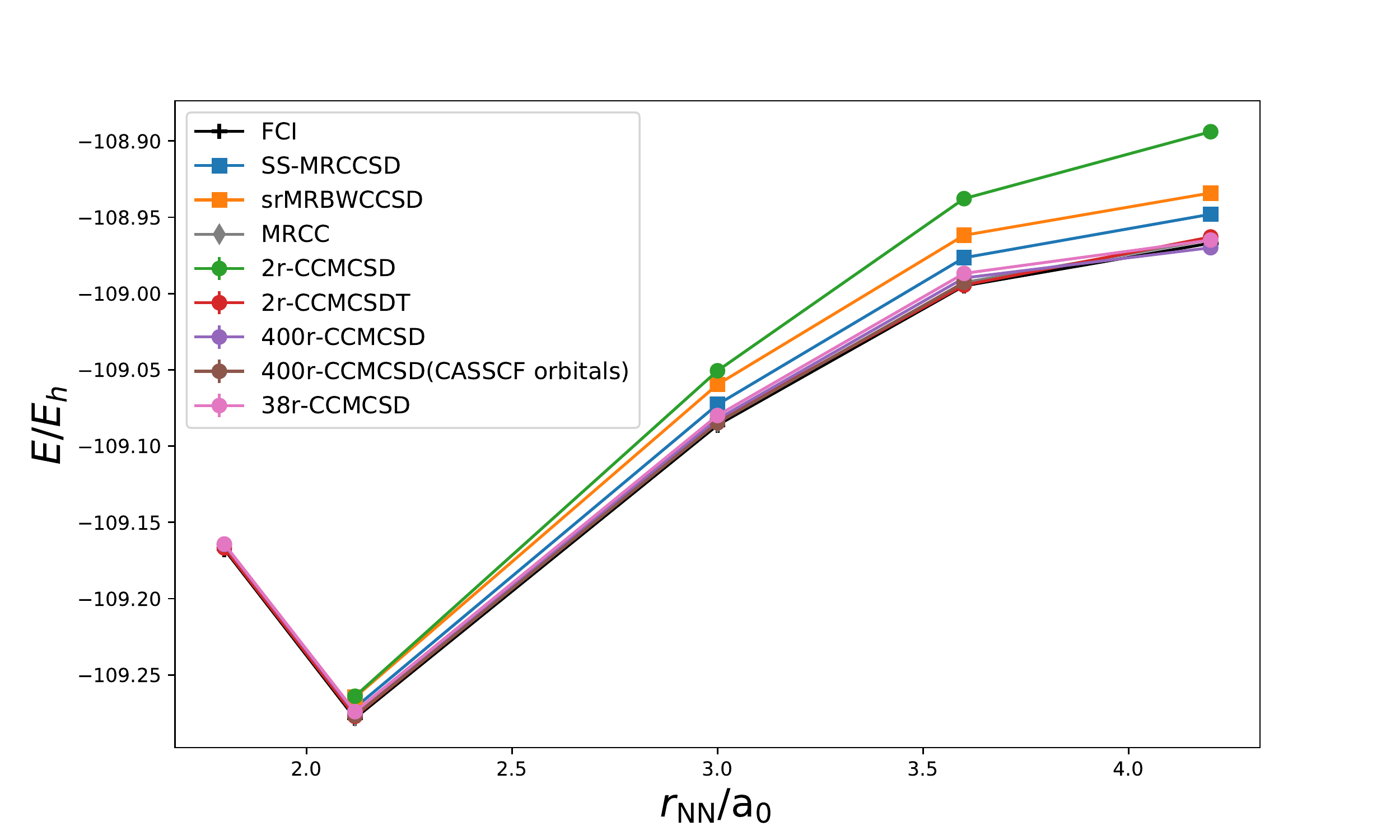}}\\
\subfloat {\includegraphics[width=0.5\textwidth, center]{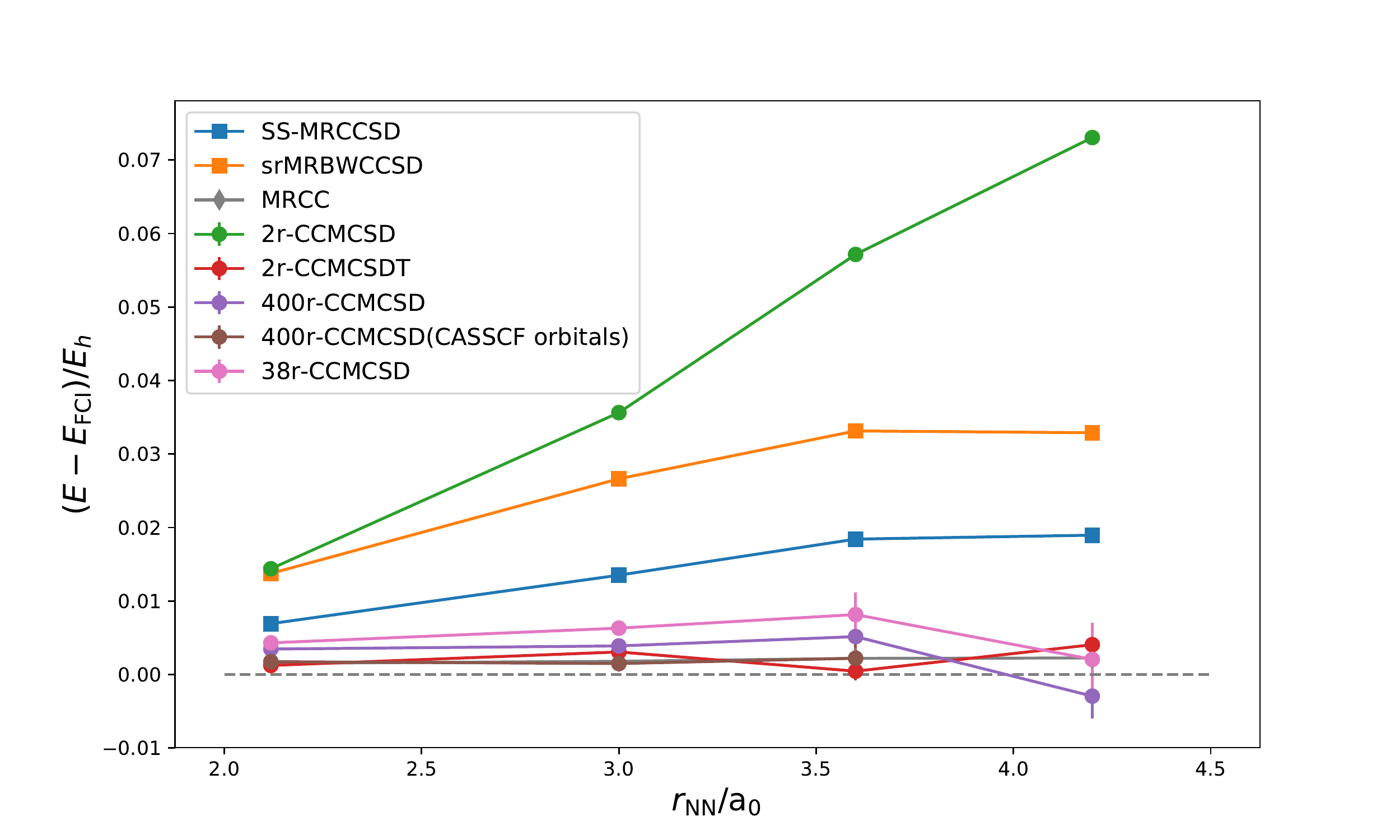}}
\caption {\protect\raggedright \footnotesize  Comparison of mrCC and mr-CCMC methods to FCI along
the N$_2$ binding curve in the Dunning cc-pVDZ basis set. Conventional multireference 
results are from Ref. \citenum{Chan2004} and \citenum{Das2010}. While 2r-CCMCSD 
(based on either canonical HF or CASSCF orbitals) underperforms relative
to these methods, accuracy is satisfactorily regained by increasing the number of references or
moving to 2r-CCMCSDT.}
\label{fig:binding2}
\end{figure}

Figure \ref{fig:binding2} shows the difference between various implementations of MRCC, 
at CCSD level\cite{Das2010, Chan2004} and the FCI energy along the N$_2$ binding curve. It
can be easily observed that 2r\nobreakdash-CCMCSDT performs as well as the best of these
methods, while 2r\nobreakdash-CCMCSD shows a significant deviation from the FCI values. We 
believe that the primary cause of this is the fact that all conventional 
methods are built on top of a CASSCF calculation in the N$_2$ (6,6) 
CAS,\cite{Das2010} with all double excitations out of this CAS considered. It 
is immediately obvious that the spanned space of such calculations is a large 
superset of the space our two-reference CCSD calculation spans, which could be
expected to improve the accuracy of these calculations.

To obtain a fairer comparison, we have included all 400 determinants in the 
CAS as references in our calculation (400r\nobreakdash-CCSD), allowing double
excitations out of each. Including the CAS in such a way is equivalent to using a CASCI
reference wavefunction and significantly improves
the quality of the obtained correlation energy (see Figure \ref{fig:binding2}), 
yielding a method that outperforms all but the most accurate conventional methods. The remaining 
gap can be bridged by using CASSCF rather than HF orbitals, however this comes at an increased computational cost.

\subsubsection{The H$_2$O molecule}
\begin{figure}[h]
\centering
\subfloat {\includegraphics[width=0.5\textwidth]{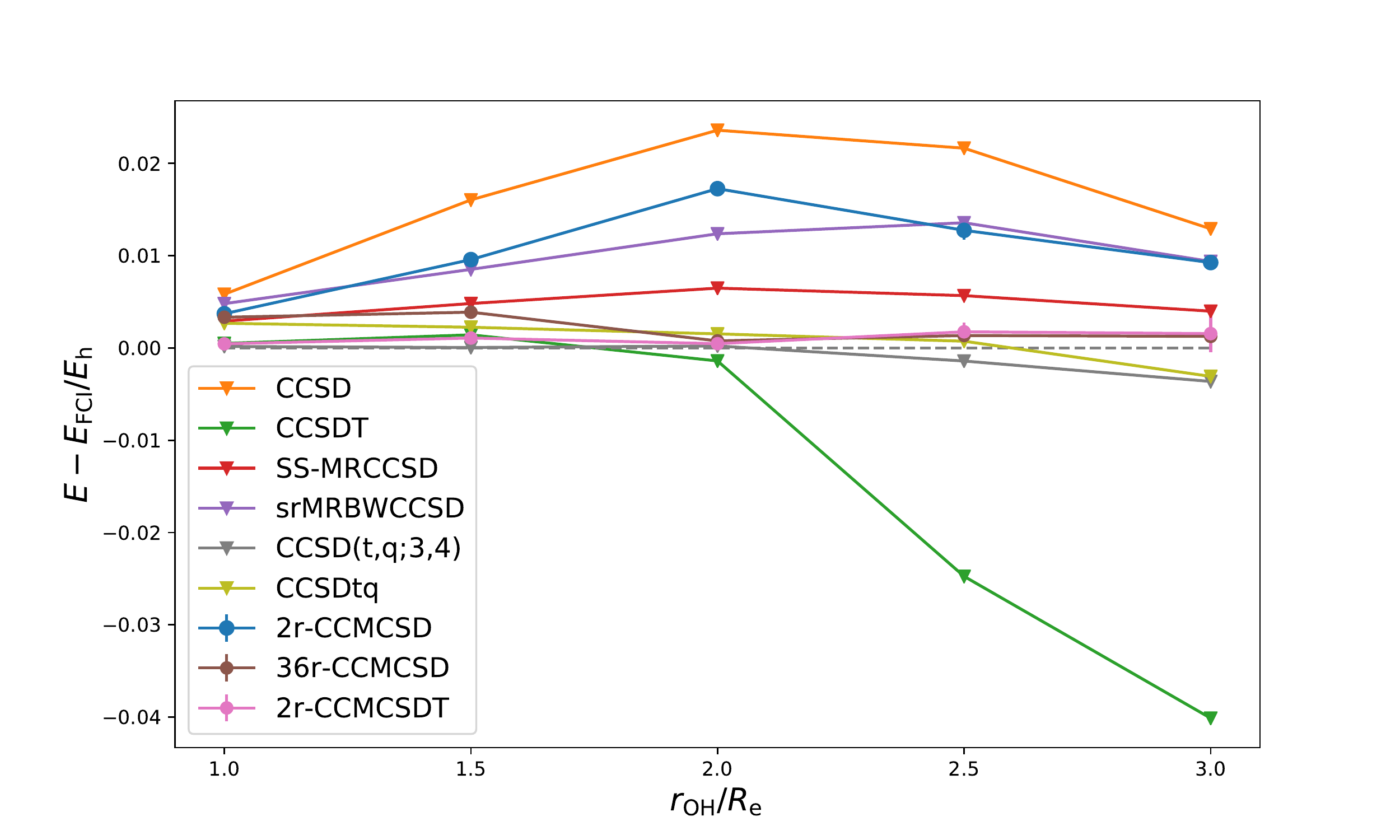}}
\caption{\protect\raggedright \footnotesize Coupled cluster energies of H$_2$O relative to the FCI energy, during
the symmetric stretch from $R_\mathrm{OH} = R_\mathrm{e}$ to $R_\mathrm{OH} = 3R_\mathrm{e}$, with
the angle fixed at 110.6$\deg$.} 
\label{fig:water}
\end{figure}

We also investigate the symmetric dissociation of the water molecule, over a range of
OH bond lengths raging from the equilibrium value $R_\mathrm{e} = 1.84345 a_0$ to $3R_\mathrm{e}$,
with the HOH angle fixed at $110.6\deg$. As can be seen from Figure \ref{fig:water},
for this system the CCSDT description fails at long bond lengths. By comparison, CCSD\footnote{The performance of
all is highly dependent of the exact Hartree--Fock reference used at $r_\mathrm{OH} = 3 R_\mathrm{e}$, where there
are two low lying RHF states. One, with $E = -75.34439 Hartree$, gives the results shown in Figure \ref{fig:water} 
for CCSD, CCSDT and mr-CCMC. The other, with $E = -75.4341998$ Hartree, causes both conventional CCSD and CCMCSD, 
CCMCSDT and mr-CCMCSDT to converge to a metastable excited state with $E_\mathrm{corr} \approx -0.4$ Hartree.} 
continues to provide reasonable descriptions 
across the binding curve. As in the case of N$_2$, this molecule has been studied using 
state-specific MRCC methods,\cite{Das2010} based  on the (4,4) CASSCF wavefunction as 
a reference. Both SS-MRCCSD and sr-MRBWCCSD consistently give errors of less than 5 and 15 
miliHartree respectively, relative to the FCI results. The CCSDtq method has also been applied 
to this system, giving errors consistently below 3 miliHartree, which can be reduced by applying 
further corrections.\cite{Piecuch1999, Bauman2017}
 
2r-CCMCSD, using the highest excited determinant in the (4,4) CAS as a secondary reference performs 
comparably to sr-MRBWCCSD, however 2r-CCMCSDT shows a significant improvement, with errors of less than 1.5 mHartrees 
across the entire binding curve. Unlike its single-reference counterpart, 2r-CCSDT provides a consistent
description of the system at all bond lengths.
As before, we can use all determinants in the (4,4) CAS as references for a CCMCSD method. Once again,
we observe a significant improvement in the quality of our estimates, generally outperforming conventional
MRCC methods, but not 2r-CCMCSDT or CCSDtq.

\subsection{Bridging the gap}
While achieving results of similar quality for N$_2$, it is worth noting that the stochastic Hilbert space
of 2r\nobreakdash-CCMCSDT is less than half of that of 400r-CCMCSD (68000 vs. 151100 
determinants). A direct comparison of these Hilbert spaces shows that, rather 
than the 2r\nobreakdash-CCMCSDT calculation spanning a strict subset of the 400r-CCMCSD space, 
they only partially overlap. The CAS shows significant redundancy in spanning this overlap,
with an average of 9 CAS determinants connected to any
(connected) determinants. However, there are determinants in the overlap that are solely
connected to one CAS determinant. Altogether these connect to only 38 of the CAS determinants
and it turns out these 38 determinants are also sufficient to span the whole overlap. This suggests 
that the significant part of the wavefunction is encoded in this subspace. The flexibility
mr-CCMC has in terms of defining references and their accepted cluster excitation levels
allows us to easily investigate this hypothesis. Indeed, an mr-CCSD calculation using 
these 38 determinants as references recovers $98.7\%$ of the correlation energy at
$r=3.6 a_0$, while decreasing the Hilbert space (and therefore memory cost) by 
$82\%$ compared to the 400r-CCSD case. It maintains this level of accuracy consistently 
across the binding curve, as can be seen in Figure \ref{fig:binding2}

\begin{figure}[h]
\centering
\subfloat {\includegraphics[width=0.5\textwidth]{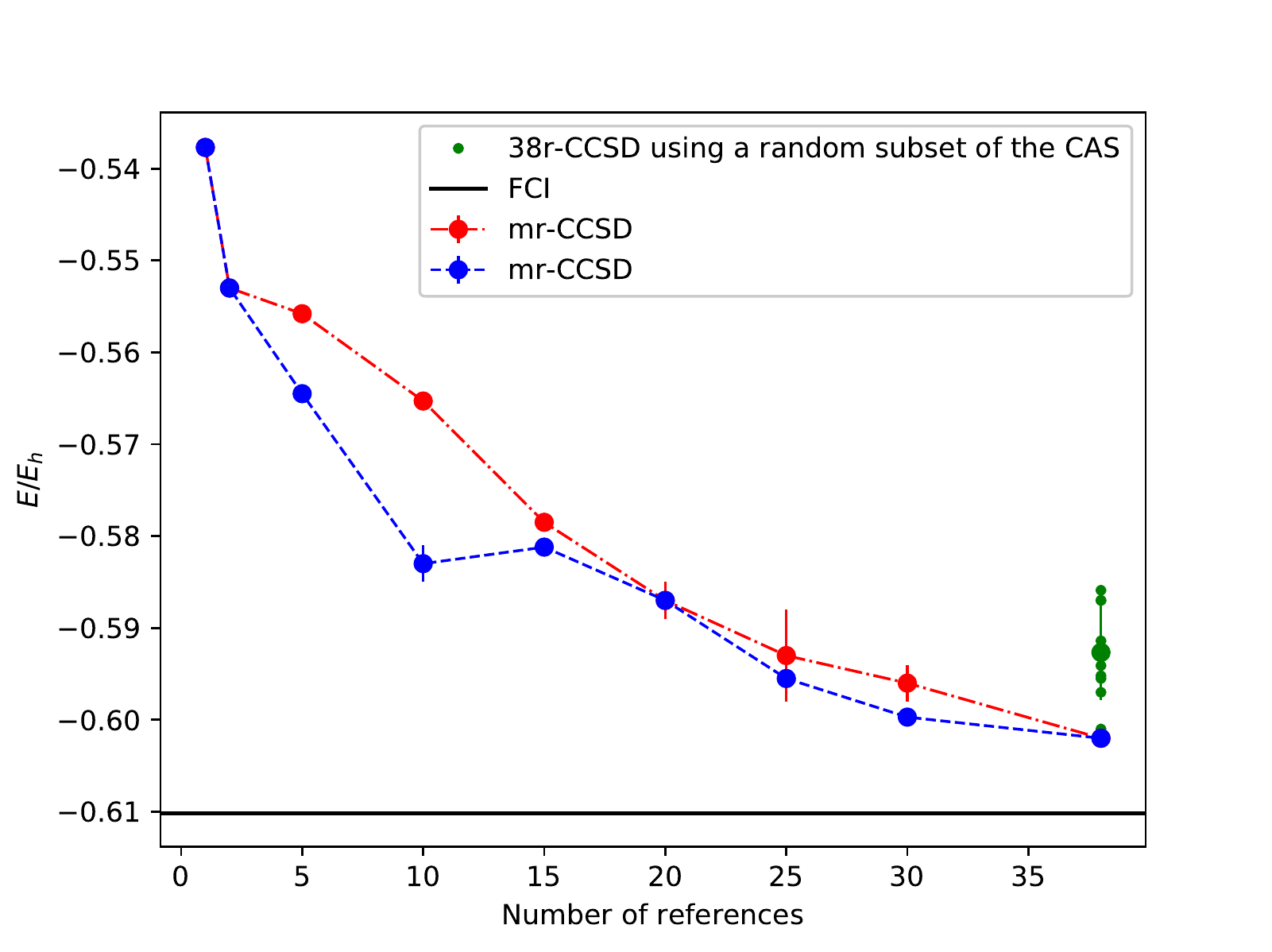}}
\caption{\protect\raggedright\footnotesize Convergence of the mr-CCMCSD energy with increasing number of 
references. The red and blue curves correspond to different orders of inclusion of references from the same 38-determinant set.}
\label{fig:multiref_conv}
\end{figure}

\begin{figure}[h]
\centering
\subfloat {\includegraphics[width=0.5\textwidth]{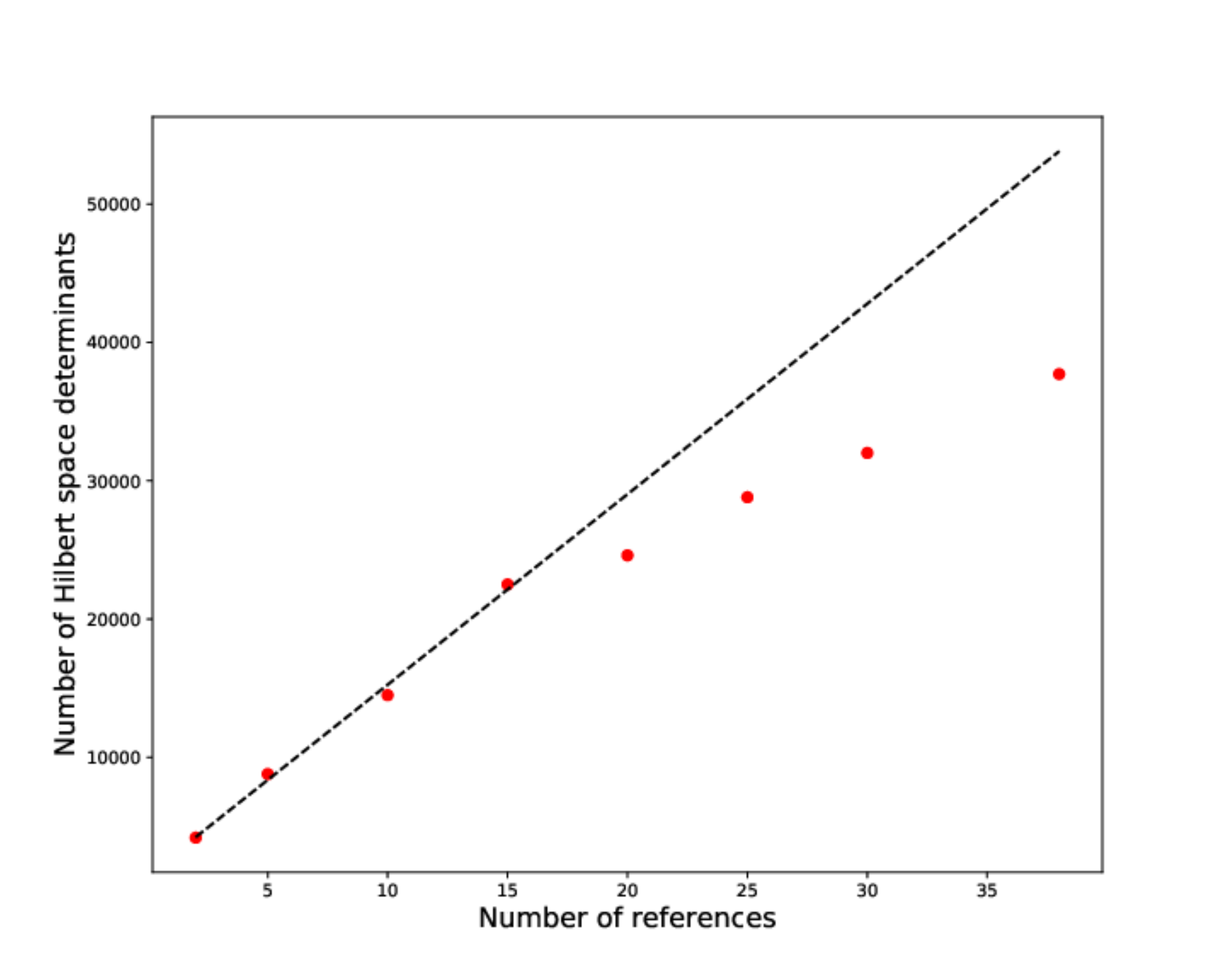}}
\caption{\protect\raggedright \footnotesize Variation of the size of the Hilbert space of a mr-CCMCSD calculation with 
the number of references for cc-pVDZ N$_2$. As expected, the size of the Hilbert space increases
sub-linearly with the number of references.} 
\label{fig:multiref_mem}
\end{figure}

The mr-CCMC method shows fast convergence of the correlation energy with increasing number of
references from this subset (see Figure \ref{fig:multiref_conv}). While the exact details
of the convergence depend on the order in which the references are included, the behaviour
is significantly outside the standard deviation of a randomly selected set of 38 references,
supporting the idea that these references and their excitations encode 
the significant part of the wavefunction. We also observe the expected sub-linear scaling of
memory cost with number of references, as their spawned spaces begin to overlap (see Figure 
\ref{fig:multiref_mem}).

In order to obtain this optimised reference
space, one requires knowledge of the spanned Hilbert spaces of the larger 400r-CCSD 
and 2r-CCSDT calculations. In this work, the information was acquired 
from stochastic snapshots of the two calculations, however a list of all determinants 
in the Hilbert space of each calculation can be easily generated if the references and 
excitation levels of the methods are known. This
could then be analysed in the same way we have done here and used to predict an
optimised, less computationally expensive method, without incurring the cost of actually running the
more demanding calculations.

\section{Conclusions}
We have successfully implemented a simple multireference technique within 
the framework of stochastic coupled cluster. The method shows a systematic 
improvement over single-reference CCMC, giving high-accuracy energy 
estimates in known strongly correlated molecular systems. The memory requirements are 
expected to scale sublinearly with the number of references used. This 
scaling is significantly better than the one expected with increasing the 
truncation level in a large Hilbert space, making the technique likely 
useful for the treatment of more complicated systems, with multiple highly 
weighted determinants in the true ground state. Significantly, in most cases the performs
at least as well as many deterministic multireference methods, while providing
a simple algorithm and significant possibilities for expansion. 

In one case, we have shown that lowered accuracy may be correlated to
the absence of some potentially significant clusters from
our expansion. The effect of including these clusters will be investigated further.
We have also observed that the choice of one electron orbitals can affect the quality
of the mr-CCMC results. We are interested in investigating iterative schemes to optimise
the orbitals used.

The method also shows great flexibility in the choice of reference and excitation space used, 
without significant effects on the stability and general behaviour of the calculations. 
This allows for potential detailed investigation into the structure of coupled cluster wavefunctions,
as well as potential optimised computations, using the minimal required reference space.

The extent to which the use of multiple references improves the correlation 
energy is system dependent, which may be at least partly due to the 
different quality of the secondary references. Therefore, a systematic way 
of selecting the best secondary references, especially in systems where 
chemical intuition is lacking, is of further interest. This could potentially be done
by iteratively modifying the reference space, using an amplitude threshold,
similarly to what is done in the initiator approach or selected CI.
\cite{Huron1973, Evangelisti1983, Tubman2016, Holmes2016a, Schriber2016} A connectivity
criterion could also be implemented. With this refinement, 
we expect that this formulation of stochastic multireference coupled cluster
could provide a flexible and robust method to compute accurate energies for a wide range 
of strongly correlated systems.

\begin{acknowledgement}
M-A.F. is grateful to Magdalene College, Cambridge for summer project 
funding and to the Cambridge Trust and Corpus Christi College for a
studentship. C.J.C.S. is grateful to the Sims Fund for a studentship
and A.J.W.T. to the Royal Society for a University Research Fellowship 
under Grant No. UF160398. All are grateful for support under ARCHER
Leadership Project grant e507.

Molecular orbital integrals were generated using PySCF\cite{pyscf}, 
Psi4\cite{psi4} and Q-Chem\cite{QCHEM41}. CASSCF orbitals were obtained using PySCF or ORCA\cite{ORCA}. Stochastic post-Hartree Fock and 
some FCI calculations were performed using a development version of HANDE-QMC
.\cite{HANDE2018} Deterministic CC calculations for N$_2$ at 1.8$a_0$ were 
performed in MRCC.\cite{mrcc}
\end{acknowledgement}

\begin{suppinfo}
Numerical values for CCMC energies, as well as Hartree--Fock and FCI references.
\end{suppinfo}

\bibliography{./biblio}

\providecommand{\latin}[1]{#1}
\providecommand*\mcitethebibliography{\thebibliography}
\csname @ifundefined\endcsname{endmcitethebibliography}
  {\let\endmcitethebibliography\endthebibliography}{}
\begin{mcitethebibliography}{82}
\providecommand*\natexlab[1]{#1}
\providecommand*\mciteSetBstSublistMode[1]{}
\providecommand*\mciteSetBstMaxWidthForm[2]{}
\providecommand*\mciteBstWouldAddEndPuncttrue
  {\def\EndOfBibitem{\unskip.}}
\providecommand*\mciteBstWouldAddEndPunctfalse
  {\let\EndOfBibitem\relax}
\providecommand*\mciteSetBstMidEndSepPunct[3]{}
\providecommand*\mciteSetBstSublistLabelBeginEnd[3]{}
\providecommand*\EndOfBibitem{}
\mciteSetBstSublistMode{f}
\mciteSetBstMaxWidthForm{subitem}{(\alph{mcitesubitemcount})}
\mciteSetBstSublistLabelBeginEnd
  {\mcitemaxwidthsubitemform\space}
  {\relax}
  {\relax}

\bibitem[Lyakh \latin{et~al.}(2011)Lyakh, Musia{\l}, Lotrich, and
  Bartlett]{Lyakh2011}
Lyakh,~D.~I.; Musia{\l},~M.; Lotrich,~V.~F.; Bartlett,~R.~J. {Multireference
  Nature of Chemistry: The Coupled-Cluster View}. \emph{Chem. Rev.}
  \textbf{2011}, 182\relax
\mciteBstWouldAddEndPuncttrue
\mciteSetBstMidEndSepPunct{\mcitedefaultmidpunct}
{\mcitedefaultendpunct}{\mcitedefaultseppunct}\relax
\EndOfBibitem
\bibitem[{\v{C}}{\'{i}}{\v{z}}ek(1966)]{Cizek1966}
{\v{C}}{\'{i}}{\v{z}}ek,~J. {On the Correlation Problem in Atomic and Molecular
  Systems. Calculation of Wavefunction Components in Ursell-Type Expansion
  Using Quantum-Field Theoretical Methods}. \emph{J. Chem. Phys} \textbf{1966},
  \emph{45}, 4256\relax
\mciteBstWouldAddEndPuncttrue
\mciteSetBstMidEndSepPunct{\mcitedefaultmidpunct}
{\mcitedefaultendpunct}{\mcitedefaultseppunct}\relax
\EndOfBibitem
\bibitem[{\v{C}}{\'{i}}{\v{z}}ek(1969)]{Cizek1969}
{\v{C}}{\'{i}}{\v{z}}ek,~J. On the Use of the Cluster Expansion and the
  Technique of Diagrams in Calculations of Correlation Effects in Atoms and
  Molecules. \emph{Adv. Chem. Phys.} \textbf{1969}, \emph{24}, 35\relax
\mciteBstWouldAddEndPuncttrue
\mciteSetBstMidEndSepPunct{\mcitedefaultmidpunct}
{\mcitedefaultendpunct}{\mcitedefaultseppunct}\relax
\EndOfBibitem
\bibitem[Chan \latin{et~al.}(2004)Chan, K{\'{a}}llay, and Gauss]{Chan2004}
Chan,~G. K.-L.; K{\'{a}}llay,~M.; Gauss,~J. State-of-the-art density matrix
  renormalization group and coupled cluster theory studies of the nitrogen
  binding curve. \emph{J. Chem. Phys.} \textbf{2004}, \emph{121}, 6110\relax
\mciteBstWouldAddEndPuncttrue
\mciteSetBstMidEndSepPunct{\mcitedefaultmidpunct}
{\mcitedefaultendpunct}{\mcitedefaultseppunct}\relax
\EndOfBibitem
\bibitem[Jeziorski and Monkhorst(1981)Jeziorski, and Monkhorst]{Jeziorski1981}
Jeziorski,~B.; Monkhorst,~H.~J. {Coupled-cluster method for multideterminantal
  reference states}. \emph{Phys. Rev. A} \textbf{1981}, \emph{24}, 1668\relax
\mciteBstWouldAddEndPuncttrue
\mciteSetBstMidEndSepPunct{\mcitedefaultmidpunct}
{\mcitedefaultendpunct}{\mcitedefaultseppunct}\relax
\EndOfBibitem
\bibitem[Piecuch and Paldus(1992)Piecuch, and Paldus]{Piecuch1992}
Piecuch,~P.; Paldus,~J. {Orthogonally spin-adapted multi-reference Hilbert
  space coupled-cluster formalism: diagrammatic formulation*}. \emph{Theor.
  Chim. Acta} \textbf{1992}, \emph{83}, 69\relax
\mciteBstWouldAddEndPuncttrue
\mciteSetBstMidEndSepPunct{\mcitedefaultmidpunct}
{\mcitedefaultendpunct}{\mcitedefaultseppunct}\relax
\EndOfBibitem
\bibitem[Piecuch and Paldus(1994)Piecuch, and Paldus]{Piecuch1994a}
Piecuch,~P.; Paldus,~J. {Orthogonally spin-adapted state-universal
  coupled-cluster formalism: Implementation of the complete two-reference
  theory including cubic and quartic coupling terms}. \emph{J. Chem. Phys}
  \textbf{1994}, \emph{101}, 5875\relax
\mciteBstWouldAddEndPuncttrue
\mciteSetBstMidEndSepPunct{\mcitedefaultmidpunct}
{\mcitedefaultendpunct}{\mcitedefaultseppunct}\relax
\EndOfBibitem
\bibitem[Mahapatra \latin{et~al.}(1998)Mahapatra, Datta, and
  Mukherjee]{Mahapatra1998}
Mahapatra,~U.~S.; Datta,~B.; Mukherjee,~D. {A state-specific multi-reference
  coupled cluster formalism with molecular applications}. \emph{Mol. Phys.}
  \textbf{1998}, \emph{94}, 157\relax
\mciteBstWouldAddEndPuncttrue
\mciteSetBstMidEndSepPunct{\mcitedefaultmidpunct}
{\mcitedefaultendpunct}{\mcitedefaultseppunct}\relax
\EndOfBibitem
\bibitem[Mahapatra \latin{et~al.}(1999)Mahapatra, Datta, and
  Mukherjee]{Mahapatra1999}
Mahapatra,~U.~S.; Datta,~B.; Mukherjee,~D. {A Size-Consistent State-Specific
  Multireference Coupled Cluster Theory: Formal Developments and Molecular
  Applications}. \emph{J. Chem. Phys.} \textbf{1999}, \emph{110}, 6171\relax
\mciteBstWouldAddEndPuncttrue
\mciteSetBstMidEndSepPunct{\mcitedefaultmidpunct}
{\mcitedefaultendpunct}{\mcitedefaultseppunct}\relax
\EndOfBibitem
\bibitem[Kowalski and Piecuch(2001)Kowalski, and Piecuch]{Kowalski2001a}
Kowalski,~K.; Piecuch,~P. {Extension of the method of moments of
  coupled-cluster equations to a multireference wave operator formalism}.
  \emph{J. Mol. Struct. THEOCHEM} \textbf{2001}, \emph{547}, 191\relax
\mciteBstWouldAddEndPuncttrue
\mciteSetBstMidEndSepPunct{\mcitedefaultmidpunct}
{\mcitedefaultendpunct}{\mcitedefaultseppunct}\relax
\EndOfBibitem
\bibitem[Chattopadhyay \latin{et~al.}(2004)Chattopadhyay, Pahari, Mukherjee,
  and Mahapatra]{Chattopadhyay2004}
Chattopadhyay,~S.; Pahari,~D.; Mukherjee,~D.; Mahapatra,~U.~S. A state-specific
  approach to multireference coupled electron-pair approximation like methods:
  Development and applications. \emph{J. Chem. Phys.} \textbf{2004},
  \emph{120}, 5968\relax
\mciteBstWouldAddEndPuncttrue
\mciteSetBstMidEndSepPunct{\mcitedefaultmidpunct}
{\mcitedefaultendpunct}{\mcitedefaultseppunct}\relax
\EndOfBibitem
\bibitem[Hanrath(2005)]{Hanrath2005}
Hanrath,~M. An exponential multireference wave-function Ansatz. \emph{J. Chem.
  Phys.} \textbf{2005}, \emph{123}, 084102\relax
\mciteBstWouldAddEndPuncttrue
\mciteSetBstMidEndSepPunct{\mcitedefaultmidpunct}
{\mcitedefaultendpunct}{\mcitedefaultseppunct}\relax
\EndOfBibitem
\bibitem[Mukherjee \latin{et~al.}(1975)Mukherjee, {Kumar Moitra}, and
  Mukhopadhyay]{Mukherjee1975}
Mukherjee,~D.; {Kumar Moitra},~R.; Mukhopadhyay,~A. Correlation problem in
  open-shell atoms and molecules. \emph{Mol. Phys.} \textbf{1975}, \emph{30},
  1861\relax
\mciteBstWouldAddEndPuncttrue
\mciteSetBstMidEndSepPunct{\mcitedefaultmidpunct}
{\mcitedefaultendpunct}{\mcitedefaultseppunct}\relax
\EndOfBibitem
\bibitem[Oliphant and Adamowicz(1991)Oliphant, and Adamowicz]{Oliphant1991}
Oliphant,~N.; Adamowicz,~L. {Multireference coupled-cluster method using a
  single-reference formalism}. \emph{J. Chem. Phys.} \textbf{1991}, \emph{94},
  1229\relax
\mciteBstWouldAddEndPuncttrue
\mciteSetBstMidEndSepPunct{\mcitedefaultmidpunct}
{\mcitedefaultendpunct}{\mcitedefaultseppunct}\relax
\EndOfBibitem
\bibitem[Piecuch \latin{et~al.}(1993)Piecuch, Oliphant, and
  Adamowicz]{Piecuch1993a}
Piecuch,~P.; Oliphant,~N.; Adamowicz,~L. {A state-selective multireference
  coupled-cluster theory employing the single-reference formalism}. \emph{J.
  Chem. Phys} \textbf{1993}, \emph{99}, 1875\relax
\mciteBstWouldAddEndPuncttrue
\mciteSetBstMidEndSepPunct{\mcitedefaultmidpunct}
{\mcitedefaultendpunct}{\mcitedefaultseppunct}\relax
\EndOfBibitem
\bibitem[Piecuch \latin{et~al.}(1999)Piecuch, Kucharski, and
  Bartlett]{Piecuch1999}
Piecuch,~P.; Kucharski,~S.~A.; Bartlett,~R.~J. Coupled-cluster methods with
  internal and semi-internal triply and quadruply excited clusters: CCSDt and
  CCSDtq approaches. \emph{J. Chem. Phys.} \textbf{1999}, \emph{110},
  1875\relax
\mciteBstWouldAddEndPuncttrue
\mciteSetBstMidEndSepPunct{\mcitedefaultmidpunct}
{\mcitedefaultendpunct}{\mcitedefaultseppunct}\relax
\EndOfBibitem
\bibitem[Kowalski and Piecuch(2000)Kowalski, and Piecuch]{Kowalski2000}
Kowalski,~K.; Piecuch,~P. {The active-space equation-of-motion coupled-cluster
  methods for excited electronic states: The EOMCCSDt approach}. \emph{J. Chem.
  Phys.} \textbf{2000}, \emph{113}, 1715\relax
\mciteBstWouldAddEndPuncttrue
\mciteSetBstMidEndSepPunct{\mcitedefaultmidpunct}
{\mcitedefaultendpunct}{\mcitedefaultseppunct}\relax
\EndOfBibitem
\bibitem[Lyakh \latin{et~al.}(2005)Lyakh, Ivanov, and Adamowicz]{Lyakh2005}
Lyakh,~D.~I.; Ivanov,~V.~V.; Adamowicz,~L. Automated generation of
  coupled-cluster diagrams: Implementation in the multireference state-specific
  coupled-cluster approach with the complete-active-space reference. \emph{J.
  Chem. Phys.} \textbf{2005}, \emph{122}, 024108\relax
\mciteBstWouldAddEndPuncttrue
\mciteSetBstMidEndSepPunct{\mcitedefaultmidpunct}
{\mcitedefaultendpunct}{\mcitedefaultseppunct}\relax
\EndOfBibitem
\bibitem[Yanai and Chan(2006)Yanai, and Chan]{Yanai2006}
Yanai,~T.; Chan,~G. K.-L. Canonical transformation theory for multireference
  problems. \emph{J. Chem. Phys.} \textbf{2006}, \emph{124}, 194106\relax
\mciteBstWouldAddEndPuncttrue
\mciteSetBstMidEndSepPunct{\mcitedefaultmidpunct}
{\mcitedefaultendpunct}{\mcitedefaultseppunct}\relax
\EndOfBibitem
\bibitem[Fang and Li(2007)Fang, and Li]{Fang2007}
Fang,~T.; Li,~S. Block correlated coupled cluster theory with a complete
  active-space self-consistent-field reference function: The formulation and
  test applications for single bond breaking. \emph{J. Chem. Phys.}
  \textbf{2007}, \emph{127}, 204108\relax
\mciteBstWouldAddEndPuncttrue
\mciteSetBstMidEndSepPunct{\mcitedefaultmidpunct}
{\mcitedefaultendpunct}{\mcitedefaultseppunct}\relax
\EndOfBibitem
\bibitem[Li and Paldus(2008)Li, and Paldus]{Li2008}
Li,~X.; Paldus,~J. Partially linearized, fully size-extensive, and reduced
  multireference coupled-cluster methods. I. Formalism and mutual relationship.
  \emph{J. Chem. Phys.} \textbf{2008}, \emph{128}, 144118\relax
\mciteBstWouldAddEndPuncttrue
\mciteSetBstMidEndSepPunct{\mcitedefaultmidpunct}
{\mcitedefaultendpunct}{\mcitedefaultseppunct}\relax
\EndOfBibitem
\bibitem[Neuscamman \latin{et~al.}(2010)Neuscamman, Yanai, and
  Chan]{Neuscamman2010}
Neuscamman,~E.; Yanai,~T.; Chan,~G. K.-L. {Strongly contracted canonical
  transformation theory}. \emph{J. Chem. Phys.} \textbf{2010}, \emph{132},
  024106\relax
\mciteBstWouldAddEndPuncttrue
\mciteSetBstMidEndSepPunct{\mcitedefaultmidpunct}
{\mcitedefaultendpunct}{\mcitedefaultseppunct}\relax
\EndOfBibitem
\bibitem[Evangelista and Gauss(2011)Evangelista, and Gauss]{Evangelista2011}
Evangelista,~F.~A.; Gauss,~J. An orbital-invariant internally contracted
  multireference coupled cluster approach. \emph{J. Chem. Phys.} \textbf{2011},
  \emph{134}, 114102\relax
\mciteBstWouldAddEndPuncttrue
\mciteSetBstMidEndSepPunct{\mcitedefaultmidpunct}
{\mcitedefaultendpunct}{\mcitedefaultseppunct}\relax
\EndOfBibitem
\bibitem[Shen and Piecuch(2012)Shen, and Piecuch]{Shen2012}
Shen,~J.; Piecuch,~P. {Combining active-space coupled-cluster methods with
  moment energy corrections via the CC(P;Q) methodology, with benchmark
  calculations for biradical transition states}. \emph{J. Chem. Phys}
  \textbf{2012}, \emph{136}, 144104\relax
\mciteBstWouldAddEndPuncttrue
\mciteSetBstMidEndSepPunct{\mcitedefaultmidpunct}
{\mcitedefaultendpunct}{\mcitedefaultseppunct}\relax
\EndOfBibitem
\bibitem[Mukherjee \latin{et~al.}(1977)Mukherjee, {Kumar Moitra}, and
  Mukhopadhyay]{Mukherjee1977}
Mukherjee,~D.; {Kumar Moitra},~R.; Mukhopadhyay,~A. Applications of a
  non-perturbative many-body formalism to general open-shell atomic and
  molecular problems: calculation of the ground and the lowest $\pi$-$\pi$*
  singlet and triplet energies and the first ionization potential of
  trans-butadiene. \emph{Mol. Phys.} \textbf{1977}, \emph{33}, 955\relax
\mciteBstWouldAddEndPuncttrue
\mciteSetBstMidEndSepPunct{\mcitedefaultmidpunct}
{\mcitedefaultendpunct}{\mcitedefaultseppunct}\relax
\EndOfBibitem
\bibitem[Haque and Mukherjee(1984)Haque, and Mukherjee]{Haque1984}
Haque,~M.~A.; Mukherjee,~D. {Application of cluster expansion techniques to
  open-shells: Calculation of difference energies}. \emph{J. Chem. Phys.}
  \textbf{1984}, \emph{80}, 5058\relax
\mciteBstWouldAddEndPuncttrue
\mciteSetBstMidEndSepPunct{\mcitedefaultmidpunct}
{\mcitedefaultendpunct}{\mcitedefaultseppunct}\relax
\EndOfBibitem
\bibitem[Lindgren and Mukherjee(1987)Lindgren, and Mukherjee]{Lindgren1987}
Lindgren,~I.; Mukherjee,~D. {On the connectivity criteria in the open-shell
  coupled-cluster theory for general model spaces}. \emph{Phys. Rep.}
  \textbf{1987}, \emph{151}, 93\relax
\mciteBstWouldAddEndPuncttrue
\mciteSetBstMidEndSepPunct{\mcitedefaultmidpunct}
{\mcitedefaultendpunct}{\mcitedefaultseppunct}\relax
\EndOfBibitem
\bibitem[Stolarczyk and Monkhorst(1985)Stolarczyk, and
  Monkhorst]{Stolarczyk1985}
Stolarczyk,~L.~Z.; Monkhorst,~H.~J. Coupled-cluster method in Fock space. I.
  General formalism. \emph{Phys. Rev. A} \textbf{1985}, \emph{32}, 725\relax
\mciteBstWouldAddEndPuncttrue
\mciteSetBstMidEndSepPunct{\mcitedefaultmidpunct}
{\mcitedefaultendpunct}{\mcitedefaultseppunct}\relax
\EndOfBibitem
\bibitem[Kaldor(1991)]{Kaldor1991}
Kaldor,~U. {The Fock space coupled cluster method: theory and application}.
  \emph{Theor. Chim. Acta} \textbf{1991}, \emph{80}, 427\relax
\mciteBstWouldAddEndPuncttrue
\mciteSetBstMidEndSepPunct{\mcitedefaultmidpunct}
{\mcitedefaultendpunct}{\mcitedefaultseppunct}\relax
\EndOfBibitem
\bibitem[Meissner(1996)]{Meissner1996}
Meissner,~L. {On multiple solutions of the Fock-space coupled-cluster method}.
  \emph{Chem. Phys. Lett.} \textbf{1996}, \emph{255}, 244\relax
\mciteBstWouldAddEndPuncttrue
\mciteSetBstMidEndSepPunct{\mcitedefaultmidpunct}
{\mcitedefaultendpunct}{\mcitedefaultseppunct}\relax
\EndOfBibitem
\bibitem[Meissner(1998)]{Meissner1998}
Meissner,~L. {Fock-space coupled-cluster method in the intermediate Hamiltonian
  formulation: Model with singles and doubles}. \emph{J. Chem. Phys.}
  \textbf{1998}, \emph{108}, 9227\relax
\mciteBstWouldAddEndPuncttrue
\mciteSetBstMidEndSepPunct{\mcitedefaultmidpunct}
{\mcitedefaultendpunct}{\mcitedefaultseppunct}\relax
\EndOfBibitem
\bibitem[Figgen \latin{et~al.}(2008)Figgen, Wedig, Stoll, Dolg, Eliav, and
  Kaldor]{Figgen2008}
Figgen,~D.; Wedig,~A.; Stoll,~H.; Dolg,~M.; Eliav,~E.; Kaldor,~U. On the
  performance of two-component energy-consistent pseudopotentials in atomic
  Fock-space coupled cluster calculations. \emph{J. Chem. Phys.} \textbf{2008},
  \emph{128}, 024106\relax
\mciteBstWouldAddEndPuncttrue
\mciteSetBstMidEndSepPunct{\mcitedefaultmidpunct}
{\mcitedefaultendpunct}{\mcitedefaultseppunct}\relax
\EndOfBibitem
\bibitem[Li and Paldus(2003)Li, and Paldus]{Li2003}
Li,~X.; Paldus,~J. {The general-model-space state-universal coupled-cluster
  method exemplified by the LiH molecule}. \emph{J. Chem. Phys.} \textbf{2003},
  \emph{119}, 5346\relax
\mciteBstWouldAddEndPuncttrue
\mciteSetBstMidEndSepPunct{\mcitedefaultmidpunct}
{\mcitedefaultendpunct}{\mcitedefaultseppunct}\relax
\EndOfBibitem
\bibitem[Balkov{\'{a}} \latin{et~al.}(1991)Balkov{\'{a}}, Kucharski, Meissner,
  and Bartlett]{Balkov1991}
Balkov{\'{a}},~A.; Kucharski,~S.~A.; Meissner,~L.; Bartlett,~R.~J. {A Hilbert
  space multi-reference coupled-cluster study of the H$_4$ model system}.
  \emph{Theor. Chim. Acta} \textbf{1991}, \emph{80}, 335\relax
\mciteBstWouldAddEndPuncttrue
\mciteSetBstMidEndSepPunct{\mcitedefaultmidpunct}
{\mcitedefaultendpunct}{\mcitedefaultseppunct}\relax
\EndOfBibitem
\bibitem[Mukhopadhyay and Mukherjee(1991)Mukhopadhyay, and
  Mukherjee]{Mukhopadhyay1991}
Mukhopadhyay,~D.; Mukherjee,~D. Molecular applications of size-extensive
  quasi-Hilbert and quasi-Fock coupled-cluster formalisms using incomplete
  model spaces. \emph{Chem. Phys. Lett.} \textbf{1991}, \emph{177}, 441\relax
\mciteBstWouldAddEndPuncttrue
\mciteSetBstMidEndSepPunct{\mcitedefaultmidpunct}
{\mcitedefaultendpunct}{\mcitedefaultseppunct}\relax
\EndOfBibitem
\bibitem[Balkov{\'{a}} and Bartlett(1994)Balkov{\'{a}}, and
  Bartlett]{Balkova1994}
Balkov{\'{a}},~A.; Bartlett,~R.~J. A multireference coupled-cluster study of
  the ground state and lowest excited states of cyclobutadiene. \emph{J. Chem.
  Phys.} \textbf{1994}, \emph{101}, 8972\relax
\mciteBstWouldAddEndPuncttrue
\mciteSetBstMidEndSepPunct{\mcitedefaultmidpunct}
{\mcitedefaultendpunct}{\mcitedefaultseppunct}\relax
\EndOfBibitem
\bibitem[Das \latin{et~al.}(2010)Das, Mukherjee, and K{\'{a}}llay]{Das2010}
Das,~S.; Mukherjee,~D.; K{\'{a}}llay,~M. Full implementation and benchmark
  studies of Mukherjee's state-specific multireference coupled-cluster ansatz.
  \emph{J. Chem. Phys.} \textbf{2010}, \emph{132}, 074103\relax
\mciteBstWouldAddEndPuncttrue
\mciteSetBstMidEndSepPunct{\mcitedefaultmidpunct}
{\mcitedefaultendpunct}{\mcitedefaultseppunct}\relax
\EndOfBibitem
\bibitem[Malrieu \latin{et~al.}(1985)Malrieu, Durand, and Daudey]{Malrieu1985}
Malrieu,~J.~P.; Durand,~P.; Daudey,~J.~P. {Intermediate Hamiltonians as a new
  class of effective Hamiltonians}. \emph{J. Phys. A} \textbf{1985}, \emph{18},
  809\relax
\mciteBstWouldAddEndPuncttrue
\mciteSetBstMidEndSepPunct{\mcitedefaultmidpunct}
{\mcitedefaultendpunct}{\mcitedefaultseppunct}\relax
\EndOfBibitem
\bibitem[Jankowski and Malinowski(1994)Jankowski, and
  Malinowski]{Jankowski1994a}
Jankowski,~K.; Malinowski,~P. A valence-universal coupled-cluster single-and
  double-excitations method for atoms. III. Solvability problems in the
  presence of intruder states. \emph{J. Phys. B} \textbf{1994}, \emph{27},
  1287\relax
\mciteBstWouldAddEndPuncttrue
\mciteSetBstMidEndSepPunct{\mcitedefaultmidpunct}
{\mcitedefaultendpunct}{\mcitedefaultseppunct}\relax
\EndOfBibitem
\bibitem[Kaldor(1988)]{Kaldor1988}
Kaldor,~U. Intruder states and incomplete model spaces in multireference
  coupled-cluster theory: The $2p^2$ states of Be. \emph{Phys. Rev. A}
  \textbf{1988}, \emph{38}, 6013\relax
\mciteBstWouldAddEndPuncttrue
\mciteSetBstMidEndSepPunct{\mcitedefaultmidpunct}
{\mcitedefaultendpunct}{\mcitedefaultseppunct}\relax
\EndOfBibitem
\bibitem[Paldus \latin{et~al.}(1993)Paldus, Piecuch, Pylypow, and
  Jeziorski]{Paldus1993}
Paldus,~J.; Piecuch,~P.; Pylypow,~L.; Jeziorski,~B. Application of
  Hilbert-space coupled-cluster theory to simple (H$_2$)$_2$ model systems:
  Planar models. \emph{Phys. Rev. A} \textbf{1993}, \emph{47}, 2738\relax
\mciteBstWouldAddEndPuncttrue
\mciteSetBstMidEndSepPunct{\mcitedefaultmidpunct}
{\mcitedefaultendpunct}{\mcitedefaultseppunct}\relax
\EndOfBibitem
\bibitem[Piecuch \latin{et~al.}(1993)Piecuch, Tobo{\l}a, and
  Paldus]{Piecuch1993}
Piecuch,~P.; Tobo{\l}a,~R.; Paldus,~J. {Approximate account of connected
  quadruply excited clusters in multi-reference Hilbert space coupled-cluster
  theory. Application to planar H$_4$ models}. \emph{Chem. Phys. Lett.}
  \textbf{1993}, \emph{210}, 243\relax
\mciteBstWouldAddEndPuncttrue
\mciteSetBstMidEndSepPunct{\mcitedefaultmidpunct}
{\mcitedefaultendpunct}{\mcitedefaultseppunct}\relax
\EndOfBibitem
\bibitem[Piecuch and Paldus(1994)Piecuch, and Paldus]{Piecuch1994}
Piecuch,~P.; Paldus,~J. {Application of Hilbert-space coupled-cluster theory to
  simple (H$_2$)$_2$ model systems. II. Nonplanar models}. \emph{Phys. Rev. A}
  \textbf{1994}, \emph{49}, 3479\relax
\mciteBstWouldAddEndPuncttrue
\mciteSetBstMidEndSepPunct{\mcitedefaultmidpunct}
{\mcitedefaultendpunct}{\mcitedefaultseppunct}\relax
\EndOfBibitem
\bibitem[K{\'{a}}llay \latin{et~al.}(2002)K{\'{a}}llay, Szalay, and
  Surj{\'{a}}n]{Kallay2002}
K{\'{a}}llay,~M.; Szalay,~P.~G.; Surj{\'{a}}n,~P.~R. A general state-selective
  multireference coupled-cluster algorithm. \emph{J. Chem. Phys.}
  \textbf{2002}, \emph{117}, 980\relax
\mciteBstWouldAddEndPuncttrue
\mciteSetBstMidEndSepPunct{\mcitedefaultmidpunct}
{\mcitedefaultendpunct}{\mcitedefaultseppunct}\relax
\EndOfBibitem
\bibitem[Nooijen \latin{et~al.}(2005)Nooijen, Shamasundar, and
  Mukherjee]{Nooijen2005}
Nooijen,~M.; Shamasundar,~K.~R.; Mukherjee,~D. Reflections on size-extensivity,
  size-consistency and generalized extensivity in many-body theory. \emph{Mol.
  Phys.} \textbf{2005}, \emph{103}, 2277\relax
\mciteBstWouldAddEndPuncttrue
\mciteSetBstMidEndSepPunct{\mcitedefaultmidpunct}
{\mcitedefaultendpunct}{\mcitedefaultseppunct}\relax
\EndOfBibitem
\bibitem[Lyakh \latin{et~al.}(2008)Lyakh, Ivanov, and Adamowicz]{Lyakh2008}
Lyakh,~D.~I.; Ivanov,~V.~V.; Adamowicz,~L. A generalization of the
  state-specific complete-active-space coupled-cluster method for calculating
  electronic excited states. \emph{J. Chem. Phys.} \textbf{2008}, \emph{128},
  074101\relax
\mciteBstWouldAddEndPuncttrue
\mciteSetBstMidEndSepPunct{\mcitedefaultmidpunct}
{\mcitedefaultendpunct}{\mcitedefaultseppunct}\relax
\EndOfBibitem
\bibitem[Booth \latin{et~al.}(2009)Booth, Thom, and Alavi]{Booth2009}
Booth,~G.~H.; Thom,~A. J.~W.; Alavi,~A. Fermion Monte Carlo without fixed
  nodes: A game of life, death, and annihilation in Slater determinant space.
  \emph{J. Chem. Phys.} \textbf{2009}, \emph{131}, 054106\relax
\mciteBstWouldAddEndPuncttrue
\mciteSetBstMidEndSepPunct{\mcitedefaultmidpunct}
{\mcitedefaultendpunct}{\mcitedefaultseppunct}\relax
\EndOfBibitem
\bibitem[Cleland \latin{et~al.}(2010)Cleland, Booth, and Alavi]{Cleland2010}
Cleland,~D.; Booth,~G.~H.; Alavi,~A. Communications: Survival of the fittest:
  Accelerating convergence in full configuration-interaction quantum Monte
  Carlo. \emph{J. Chem. Phys.} \textbf{2010}, \emph{132}, 041103\relax
\mciteBstWouldAddEndPuncttrue
\mciteSetBstMidEndSepPunct{\mcitedefaultmidpunct}
{\mcitedefaultendpunct}{\mcitedefaultseppunct}\relax
\EndOfBibitem
\bibitem[Booth \latin{et~al.}(2011)Booth, Cleland, Thom, and Alavi]{Booth2011}
Booth,~G.~H.; Cleland,~D.; Thom,~A. J.~W.; Alavi,~A. Breaking the carbon dimer:
  The challenges of multiple bond dissociation with full configuration
  interaction quantum Monte Carlo methods. \emph{J. Chem. Phys.} \textbf{2011},
  \emph{135}, 084104\relax
\mciteBstWouldAddEndPuncttrue
\mciteSetBstMidEndSepPunct{\mcitedefaultmidpunct}
{\mcitedefaultendpunct}{\mcitedefaultseppunct}\relax
\EndOfBibitem
\bibitem[Shepherd \latin{et~al.}(2012)Shepherd, Booth, Gr{\"{u}}neis, and
  Alavi]{Shepherd2012}
Shepherd,~J.~J.; Booth,~G.; Gr{\"{u}}neis,~A.; Alavi,~A. {Full configuration
  interaction perspective on the homogeneous electron gas}. \emph{Phys. Rev. B}
  \textbf{2012}, \emph{85}, 081103\relax
\mciteBstWouldAddEndPuncttrue
\mciteSetBstMidEndSepPunct{\mcitedefaultmidpunct}
{\mcitedefaultendpunct}{\mcitedefaultseppunct}\relax
\EndOfBibitem
\bibitem[Booth \latin{et~al.}(2013)Booth, Gr{\"{u}}neis, Kresse, and
  Alavi]{Booth2013}
Booth,~G.~H.; Gr{\"{u}}neis,~A.; Kresse,~G.; Alavi,~A. {Towards an exact
  description of electronic wavefunctions in real solids}. \emph{Nature}
  \textbf{2013}, \emph{493}, 365\relax
\mciteBstWouldAddEndPuncttrue
\mciteSetBstMidEndSepPunct{\mcitedefaultmidpunct}
{\mcitedefaultendpunct}{\mcitedefaultseppunct}\relax
\EndOfBibitem
\bibitem[Thom(2010)]{Thom2010}
Thom,~A. J.~W. {Stochastic Coupled Cluster Theory}. \emph{Phys. Rev. Lett.}
  \textbf{2010}, \emph{105}, 263004\relax
\mciteBstWouldAddEndPuncttrue
\mciteSetBstMidEndSepPunct{\mcitedefaultmidpunct}
{\mcitedefaultendpunct}{\mcitedefaultseppunct}\relax
\EndOfBibitem
\bibitem[Deustua \latin{et~al.}(2017)Deustua, Shen, and Piecuch]{Deustua2017}
Deustua,~J.~E.; Shen,~J.; Piecuch,~P. {Converging High-Level Coupled-Cluster
  Energetics by Monte Carlo Sampling and Moment Expansions}. \emph{Phys. Rev.
  Lett} \textbf{2017}, \emph{119}, 223003\relax
\mciteBstWouldAddEndPuncttrue
\mciteSetBstMidEndSepPunct{\mcitedefaultmidpunct}
{\mcitedefaultendpunct}{\mcitedefaultseppunct}\relax
\EndOfBibitem
\bibitem[Deustua \latin{et~al.}(2018)Deustua, Magoulas, Shen, and
  Piecuch]{Deustua2018}
Deustua,~J.~E.; Magoulas,~I.; Shen,~J.; Piecuch,~P. {Communication: Approaching
  exact quantum chemistry by cluster analysis of full configuration interaction
  quantum Monte Carlo wave functions}. \emph{J. Chem. Phys.} \textbf{2018},
  \emph{149}, 151101\relax
\mciteBstWouldAddEndPuncttrue
\mciteSetBstMidEndSepPunct{\mcitedefaultmidpunct}
{\mcitedefaultendpunct}{\mcitedefaultseppunct}\relax
\EndOfBibitem
\bibitem[Spencer and Thom(2016)Spencer, and Thom]{Spencer2016}
Spencer,~J.~S.; Thom,~A. J.~W. Developments in stochastic coupled cluster
  theory: The initiator approximation and application to the uniform electron
  gas. \emph{J. Chem. Phys.} \textbf{2016}, \emph{144}, 084108\relax
\mciteBstWouldAddEndPuncttrue
\mciteSetBstMidEndSepPunct{\mcitedefaultmidpunct}
{\mcitedefaultendpunct}{\mcitedefaultseppunct}\relax
\EndOfBibitem
\bibitem[Spencer \latin{et~al.}(2018)Spencer, Neufeld, Vigor, Franklin, and
  Thom]{Spencer2018}
Spencer,~J.~S.; Neufeld,~V.~A.; Vigor,~W.~A.; Franklin,~R. S.~T.; Thom,~A.
  J.~W. {Large scale parallelization in stochastic coupled cluster}. \emph{J.
  Chem. Phys.} \textbf{2018}, \emph{149}, 204103\relax
\mciteBstWouldAddEndPuncttrue
\mciteSetBstMidEndSepPunct{\mcitedefaultmidpunct}
{\mcitedefaultendpunct}{\mcitedefaultseppunct}\relax
\EndOfBibitem
\bibitem[Scott and Thom(2017)Scott, and Thom]{Scott2017}
Scott,~C. J.~C.; Thom,~A. J.~W. Stochastic coupled cluster theory: Efficient
  sampling of the coupled cluster expansion. \emph{J. Chem. Phys.}
  \textbf{2017}, \emph{147}, 124105\relax
\mciteBstWouldAddEndPuncttrue
\mciteSetBstMidEndSepPunct{\mcitedefaultmidpunct}
{\mcitedefaultendpunct}{\mcitedefaultseppunct}\relax
\EndOfBibitem
\bibitem[Franklin \latin{et~al.}(2016)Franklin, Spencer, Zoccante, and
  Thom]{Franklin2016}
Franklin,~R. S.~T.; Spencer,~J.~S.; Zoccante,~A.; Thom,~A. J.~W. {Linked
  coupled cluster Monte Carlo}. \emph{J. Chem. Phys.} \textbf{2016},
  \emph{144}, 044111\relax
\mciteBstWouldAddEndPuncttrue
\mciteSetBstMidEndSepPunct{\mcitedefaultmidpunct}
{\mcitedefaultendpunct}{\mcitedefaultseppunct}\relax
\EndOfBibitem
\bibitem[Holmes \latin{et~al.}(2016)Holmes, Changlani, and Umrigar]{Holmes2016}
Holmes,~A.~A.; Changlani,~H.~J.; Umrigar,~C.~J. {Efficient Heat-Bath Sampling
  in Fock Space}. \emph{J. Chem. Theory Comput.} \textbf{2016}, \emph{12},
  1561\relax
\mciteBstWouldAddEndPuncttrue
\mciteSetBstMidEndSepPunct{\mcitedefaultmidpunct}
{\mcitedefaultendpunct}{\mcitedefaultseppunct}\relax
\EndOfBibitem
\bibitem[Neufeld and Thom(2019)Neufeld, and Thom]{Neufeld2018}
Neufeld,~V.~A.; Thom,~A. J.~W. Exciting Determinants in Quantum Monte Carlo:
  Loading the Dice with Fast, Low Memory Weights. \emph{J. Chem. Theory
  Comput.} \textbf{2019}, \emph{15}, 127\relax
\mciteBstWouldAddEndPuncttrue
\mciteSetBstMidEndSepPunct{\mcitedefaultmidpunct}
{\mcitedefaultendpunct}{\mcitedefaultseppunct}\relax
\EndOfBibitem
\bibitem[Scott \latin{et~al.}(2019)Scott, {Di Remigio}, Crawford, and
  Thom]{Scott2019}
Scott,~C. J.~C.; {Di Remigio},~R.; Crawford,~T.~D.; Thom,~A. J.~W.
  {Diagrammatic Coupled Cluster Monte Carlo}. \emph{J. Phys. Chem. Lett}
  \textbf{2019}, \emph{10}, 925\relax
\mciteBstWouldAddEndPuncttrue
\mciteSetBstMidEndSepPunct{\mcitedefaultmidpunct}
{\mcitedefaultendpunct}{\mcitedefaultseppunct}\relax
\EndOfBibitem
\bibitem[Oliphant and Adamowicz(1992)Oliphant, and Adamowicz]{Oliphant1992}
Oliphant,~N.; Adamowicz,~L. The implementation of the multireference
  coupled-cluster method based on the single-reference formalism. \emph{J.
  Chem. Phys.} \textbf{1992}, \emph{96}, 3739\relax
\mciteBstWouldAddEndPuncttrue
\mciteSetBstMidEndSepPunct{\mcitedefaultmidpunct}
{\mcitedefaultendpunct}{\mcitedefaultseppunct}\relax
\EndOfBibitem
\bibitem[Piecuch and Adamowicz(1994)Piecuch, and Adamowicz]{Piecuch1994b}
Piecuch,~P.; Adamowicz,~L. {State-selective multireference coupled-cluster
  theory employing the single-reference formalism: Implementation and
  application to the H$_8$ model system}. \emph{J. Chem. Phys} \textbf{1994},
  \emph{100}, 5792\relax
\mciteBstWouldAddEndPuncttrue
\mciteSetBstMidEndSepPunct{\mcitedefaultmidpunct}
{\mcitedefaultendpunct}{\mcitedefaultseppunct}\relax
\EndOfBibitem
\bibitem[Jankowski and Paldus(1980)Jankowski, and Paldus]{Jankowski1980}
Jankowski,~K.; Paldus,~J. {Applicability of coupled-pair theories to
  quasidegenerate electronic states: A model study}. \emph{Int. J. Quantum
  Chem.} \textbf{1980}, \emph{18}, 1243\relax
\mciteBstWouldAddEndPuncttrue
\mciteSetBstMidEndSepPunct{\mcitedefaultmidpunct}
{\mcitedefaultendpunct}{\mcitedefaultseppunct}\relax
\EndOfBibitem
\bibitem[Laidig \latin{et~al.}(1987)Laidig, Saxe, and Bartlett]{Laidig1987}
Laidig,~W.~D.; Saxe,~P.; Bartlett,~R.~J. The Description of \uppercase{N}$_{2}$
  and \uppercase{F}$_{2}$ potential energy surfaces using multireference
  coupled cluster theory. \emph{J. Chem. Phys.} \textbf{1987}, \emph{86},
  887\relax
\mciteBstWouldAddEndPuncttrue
\mciteSetBstMidEndSepPunct{\mcitedefaultmidpunct}
{\mcitedefaultendpunct}{\mcitedefaultseppunct}\relax
\EndOfBibitem
\bibitem[Hehre \latin{et~al.}(1969)Hehre, Stewart, and Pople]{Hehre1969}
Hehre,~W.~J.; Stewart,~R.~F.; Pople,~J.~A. {Self-Consistent Molecular-Orbital
  Methods. I. Use of Gaussian Expansions of Slater-Type Atomic Orbitals}.
  \emph{J. Chem. Phys.} \textbf{1969}, \emph{51}, 2657\relax
\mciteBstWouldAddEndPuncttrue
\mciteSetBstMidEndSepPunct{\mcitedefaultmidpunct}
{\mcitedefaultendpunct}{\mcitedefaultseppunct}\relax
\EndOfBibitem
\bibitem[Dunning(1989)]{Dunning1989}
Dunning,~T.~H. Gaussian basis sets for use in correlated molecular
  calculations. I. The atoms boron through neon and hydrogen. \emph{J. Chem.
  Phys.} \textbf{1989}, \emph{90}, 1007\relax
\mciteBstWouldAddEndPuncttrue
\mciteSetBstMidEndSepPunct{\mcitedefaultmidpunct}
{\mcitedefaultendpunct}{\mcitedefaultseppunct}\relax
\EndOfBibitem
\bibitem[Jones(2016)]{jones2016chemistry}
Jones,~K. \emph{The Chemistry of Nitrogen: Pergamon Texts in Inorganic
  Chemistry}; Elsevier, 2016\relax
\mciteBstWouldAddEndPuncttrue
\mciteSetBstMidEndSepPunct{\mcitedefaultmidpunct}
{\mcitedefaultendpunct}{\mcitedefaultseppunct}\relax
\EndOfBibitem
\bibitem[Jankowski \latin{et~al.}(1985)Jankowski, Meissner, and
  Wasilewski]{Jankowski1985}
Jankowski,~K.; Meissner,~L.; Wasilewski,~J. {Davidson-type corrections for
  quasidegenerate states}. \emph{Int. J. Quantum Chem.} \textbf{1985},
  \emph{28}, 931--942\relax
\mciteBstWouldAddEndPuncttrue
\mciteSetBstMidEndSepPunct{\mcitedefaultmidpunct}
{\mcitedefaultendpunct}{\mcitedefaultseppunct}\relax
\EndOfBibitem
\bibitem[Bauman \latin{et~al.}(2017)Bauman, Shen, and Piecuch]{Bauman2017}
Bauman,~N.~P.; Shen,~J.; Piecuch,~P. {Combining active-space coupled-cluster
  approaches with moment energy corrections via the CC(P;Q) methodology:
  connected quadruple excitations}. \emph{Mol. Phys.} \textbf{2017},
  \emph{115}, 2860\relax
\mciteBstWouldAddEndPuncttrue
\mciteSetBstMidEndSepPunct{\mcitedefaultmidpunct}
{\mcitedefaultendpunct}{\mcitedefaultseppunct}\relax
\EndOfBibitem
\bibitem[Huron \latin{et~al.}(1973)Huron, Malrieu, and Rancurel]{Huron1973}
Huron,~B.; Malrieu,~J.~P.; Rancurel,~P. {Iterative perturbation calculations of
  ground and excited state energies from multiconfigurational zeroth-order
  wavefunctions}. \emph{J. Chem. Phys.} \textbf{1973}, \emph{58}, 5745\relax
\mciteBstWouldAddEndPuncttrue
\mciteSetBstMidEndSepPunct{\mcitedefaultmidpunct}
{\mcitedefaultendpunct}{\mcitedefaultseppunct}\relax
\EndOfBibitem
\bibitem[Evangelisti \latin{et~al.}(1983)Evangelisti, Daudey, and
  Malrieu]{Evangelisti1983}
Evangelisti,~S.; Daudey,~J.-P.; Malrieu,~J.-P. {Convergence of an improved
  CIPSI algorithm}. \emph{Chem. Phys.} \textbf{1983}, \emph{75}, 91\relax
\mciteBstWouldAddEndPuncttrue
\mciteSetBstMidEndSepPunct{\mcitedefaultmidpunct}
{\mcitedefaultendpunct}{\mcitedefaultseppunct}\relax
\EndOfBibitem
\bibitem[Tubman \latin{et~al.}(2016)Tubman, Lee, Takeshita, Head-Gordon, and
  Whaley]{Tubman2016}
Tubman,~N.~M.; Lee,~J.; Takeshita,~T.~Y.; Head-Gordon,~M.; Whaley,~K.~B. {A
  deterministic alternative to the full configuration interaction quantum Monte
  Carlo method}. \emph{J. Chem. Phys.} \textbf{2016}, \emph{145}, 044112\relax
\mciteBstWouldAddEndPuncttrue
\mciteSetBstMidEndSepPunct{\mcitedefaultmidpunct}
{\mcitedefaultendpunct}{\mcitedefaultseppunct}\relax
\EndOfBibitem
\bibitem[Holmes \latin{et~al.}(2016)Holmes, Tubman, and Umrigar]{Holmes2016a}
Holmes,~A.~A.; Tubman,~N.~M.; Umrigar,~C.~J. {Heat-Bath Configuration
  Interaction: An Efficient Selected Configuration Interaction Algorithm
  Inspired by Heat-Bath Sampling}. \emph{J. Chem. Theory Comput.}
  \textbf{2016}, \emph{12}, 3674\relax
\mciteBstWouldAddEndPuncttrue
\mciteSetBstMidEndSepPunct{\mcitedefaultmidpunct}
{\mcitedefaultendpunct}{\mcitedefaultseppunct}\relax
\EndOfBibitem
\bibitem[Schriber and Evangelista(2016)Schriber, and Evangelista]{Schriber2016}
Schriber,~J.~B.; Evangelista,~F.~A. {Communication: An adaptive configuration
  interaction approach for strongly correlated electrons with tunable
  accuracy}. \emph{J. Chem. Phys.} \textbf{2016}, \emph{144}, 161106\relax
\mciteBstWouldAddEndPuncttrue
\mciteSetBstMidEndSepPunct{\mcitedefaultmidpunct}
{\mcitedefaultendpunct}{\mcitedefaultseppunct}\relax
\EndOfBibitem
\bibitem[Sun \latin{et~al.}(2017)Sun, Berkelbach, Blunt, Booth, Guo, Li, Liu,
  McClain, Sayfutyarova, Sharma, and \latin{et al}]{pyscf}
Sun,~Q.; Berkelbach,~T.~C.; Blunt,~N.~S.; Booth,~G.~H.; Guo,~S.; Li,~Z.;
  Liu,~J.; McClain,~J.~D.; Sayfutyarova,~E.~R.; Sharma,~S.; \latin{et al},
  PySCF: the Python-based Simulations of Chemistry Framework. \emph{WIREs
  Comput. Mol. Sci.} \textbf{2017}, \emph{8}, e1340\relax
\mciteBstWouldAddEndPuncttrue
\mciteSetBstMidEndSepPunct{\mcitedefaultmidpunct}
{\mcitedefaultendpunct}{\mcitedefaultseppunct}\relax
\EndOfBibitem
\bibitem[Parrish \latin{et~al.}(2017)Parrish, Burns, Smith, Simonett, {DePrince
  III}, Hohenstein, Bozkaya, Sokolov, {Di Remigio}, Richard, Gonthier, and
  \latin{et al}]{psi4}
Parrish,~R.~M.; Burns,~L.~A.; Smith,~D. G.~A.; Simonett,~A.~C.; {DePrince
  III},~A.~E.; Hohenstein,~E.~G.; Bozkaya,~U.; Sokolov,~A.~Y.; {Di
  Remigio},~R.; Richard,~R.~M.; Gonthier,~J.~F.; \latin{et al}, Psi4 1.1: An
  Open-Source Electronic Structure Program Emphasizing Automation, Advanced
  Libraries, and Interoperability. \emph{J. Chem. Theory Comput} \textbf{2017},
  \emph{13}, 3185\relax
\mciteBstWouldAddEndPuncttrue
\mciteSetBstMidEndSepPunct{\mcitedefaultmidpunct}
{\mcitedefaultendpunct}{\mcitedefaultseppunct}\relax
\EndOfBibitem
\bibitem[Shao \latin{et~al.}(2015)Shao, Gan, Epifanovsky, Gilbert, Wormit,
  Kussmann, Lange, Behn, Deng, and Feng]{QCHEM41}
Shao,~Y.; Gan,~Z.; Epifanovsky,~E.; Gilbert,~A. T.~B.; Wormit,~M.;
  Kussmann,~J.; Lange,~A.~W.; Behn,~A.; Deng,~J.; Feng,~X.~l. Advances in
  Molecular Quantum Chemistry Contained in the Q-Chem 4 Program Package.
  \emph{Mol.\ Phys.} \textbf{2015}, \emph{113}, 184\relax
\mciteBstWouldAddEndPuncttrue
\mciteSetBstMidEndSepPunct{\mcitedefaultmidpunct}
{\mcitedefaultendpunct}{\mcitedefaultseppunct}\relax
\EndOfBibitem
\bibitem[Neese(2012)]{ORCA}
Neese,~F. {The ORCA program system}. \emph{WIREs Comput. Mol. Sci.}
  \textbf{2012}, \emph{2}, 73--78\relax
\mciteBstWouldAddEndPuncttrue
\mciteSetBstMidEndSepPunct{\mcitedefaultmidpunct}
{\mcitedefaultendpunct}{\mcitedefaultseppunct}\relax
\EndOfBibitem
\bibitem[Spencer \latin{et~al.}(2019)Spencer, Blunt, Choi, Etrych, Filip,
  Foulkes, Franklin, Handley, Malone, Neufeld, and \latin{et al}]{HANDE2018}
Spencer,~J.~S.; Blunt,~N.~S.; Choi,~S.; Etrych,~J.; Filip,~M.-A.; Foulkes,~W.
  M.~C.; Franklin,~R. S.~T.; Handley,~W.~J.; Malone,~F.~D.; Neufeld,~V.~A.;
  \latin{et al}, The HANDE-QMC Project: Open-Source Stochastic Quantum
  Chemistry from the Ground State Up. \emph{J. Chem. Theory Comput.}
  \textbf{2019}, \emph{15}, 1728\relax
\mciteBstWouldAddEndPuncttrue
\mciteSetBstMidEndSepPunct{\mcitedefaultmidpunct}
{\mcitedefaultendpunct}{\mcitedefaultseppunct}\relax
\EndOfBibitem
\bibitem[K{\'a}llay \latin{et~al.}()K{\'a}llay, Rolik, Csontos, Nagy, Samu,
  Mester, Csóka, Szabó, Ladjánszki, Szegedy, and \latin{et al}]{mrcc}
K{\'a}llay,~M.; Rolik,~Z.; Csontos,~J.; Nagy,~P.; Samu,~G.; Mester,~D.;
  Csóka,~J.; Szabó,~B.; Ladjánszki,~I.; Szegedy,~L.; \latin{et al}, MRCC, a
  Quantum Chemical Program Suite. \url{www.mrcc.hu}\relax
\mciteBstWouldAddEndPuncttrue
\mciteSetBstMidEndSepPunct{\mcitedefaultmidpunct}
{\mcitedefaultendpunct}{\mcitedefaultseppunct}\relax
\EndOfBibitem
\end{mcitethebibliography}

\end{document}